\documentclass[sigconf,screen,nonacm]{acmart}

\makeatletter

\if@ACM@authordraft
  \includecomment{draftonly}
\else
  \excludecomment{draftonly}
\fi

\newcommand{\anonymize}[2]{\if@ACM@anonymous#2\else#1\fi}
\makeatother

\usepackage[linesnumbered,ruled,vlined]{algorithm2e}
\usepackage{booktabs}
\usepackage{subcaption}
\usepackage{color,colortbl}
\usepackage{bm} 

\usepackage{enumitem}
\setlist{noitemsep}

\usepackage{cleveref}
\usepackage{todonotes}
\usepackage{tikz}

\newcommand{\falstar}{\textsc{FalStar}}
\newcommand{\aLVTS}{aLVTS}

\newcommand{\bu}{\bm{u}}
\newcommand{\by}{\bm{y}}
\newcommand{\cons}{{}} 

\newcommand{\bv}{\mathbf{v}}
\newcommand{\bb}{\mathbf{b}}
\newcommand{\bp}{\mathbf{p}}
\newcommand{\bt}{\mathbf{t}}

\newcommand{\ul}{\underline}
\newcommand{\ol}{\overline}

\newcommand{\R}{\mathbb{R}}
\newcommand{\N}{\mathbb{N}}

\newcommand{\Rn}{\R^{n}}
\newcommand{\Rm}{\R^{m}}
\newcommand{\A}{\mathcal{A}}
\newcommand{\M}{\mathcal{M}}

\newcommand{\D}{\mathcal{D}}
\renewcommand{\emptyset}{\varnothing}

\newcommand{\argmin}{\mathop{\mathrm{arg~min}}\limits}

\newcommand{\unexplored}{\mathit{unexplored}}
\newcommand{\explored}{\mathit{explored}}
\newcommand{\actions}{\A}
\newcommand{\distribution}{\D}

\newcommand{\Until}{\mathrel{\text{\upshape\sffamily U}}}

\newcommand{\AF}{\mathit{AF}}
\newcommand{\AFref}{\mathit{AF}_\text{ref}}

\setcopyright{rightsretained}
\acmConference[]{}{}{}
\acmYear{2018}

\begin{document}

\title[Fast Falsification of Hybrid Systems using Probabilistically Adaptive Input] 
      {Fast Falsification of Hybrid Systems using\\Probabilistically Adaptive Input}

\author{Gidon Ernst}
\affiliation{University of Melbourne, Australia}
\email{gidon.ernst@unimelb.edu.au}

\author{Sean Sedwards}
\affiliation{University of Waterloo, Canada}
\email{sean.sedwards@uwaterloo.ca}

\author{Zhenya Zhang}
\affiliation{National Institute of Informatics, Tokyo, Japan}
\email{zhangzy@nii.ac.jp}

\author{Ichiro Hasuo}
\affiliation{National Institute of Informatics, Tokyo, Japan}
\email{hasuo@nii.ac.jp}

\begin{abstract}
We present an algorithm that quickly finds falsifying inputs for hybrid systems,
i.e., inputs that steer the system towards violation of a given temporal logic requirement.
Our method is based on a probabilistically directed search of an increasingly fine grained
spatial and temporal discretization of the input space.
A key feature is that it adapts to the difficulty of a problem at hand,
specifically to the local complexity of each input segment, as needed for falsification.
In experiments with standard benchmarks, our approach consistently
outperforms existing techniques by a significant margin.
In recognition of the way it works and to distinguish it from previous work, we describe our method as a ``Las Vegas tree search''.
\end{abstract}



\maketitle

\begin{draftonly}
\todo[inline]{TODOs}
\begin{itemize}
\item fix commit number for \falstar
\item motivating example
      + show how the tree is constructed in a sequence of figures
\item concrete examples for~$\actions$ and~$\distribution$
\item evaluate alternative choices for~$\actions$ and~$\distribution$.
\item show some traces for the AFC violation
\end{itemize}
\end{draftonly}

\section{Introduction}
\label{sec:introduction}

The falsification problem we consider seeks a (time-bounded) input signal
that causes a hybrid system model to violate a given temporal logic specification.
A popular way to address this is to first construct a ``score function''~\cite{JegourelLegaySedwards2013} that quantifies how much of the specification has been satisfied during the course of an execution.
The falsification can then be treated as an optimization problem, which can be solved using standard algorithms.
This approach, especially using ``robustness''~\cite{FainekosPappas2009journal} as the score function, has been successfully applied, resulting in a number of now mature tools~\cite{Donze2010,Annpureddy-et-al2011}.

Despite its apparent success, the commonly used robustness semantics \cite{FainekosPappas2009journal} is in general not a perfect optimization function.
Greedy hill climbing may lead to local optima, hence robustness is only an heuristic score function~\cite{JegourelLegaySedwards2014} with respect to the falsification problem.
In practice, standard optimization algorithms overcome this limitation by including stochastic exploration.
The most sophisticated of these can also model the dynamics of the system (e.g.,~\cite{Akazaki2016}), in order to estimate the most productive direction of input signal space to explore.
There is, however, ``no free lunch''~\cite{WolpertMacready1997}, and high performance general purpose optimization algorithms are not necessarily the best choice.
For example, such algorithms often optimize with respect to the entire input trace, without exploiting the time causality of the problem, i.e., the fact that a good trace (one that eventually falsifies the property) may be dependent on a good trace prefix.

\medskip

The contribution of this paper is a fast randomized falsification algorithm (\cref{sec:algorithm}) that exploits the time-causal structure of the problem and that adapts to local complexity.
In common with alternative approaches, our algorithm searches a discretized space of input signals, but in our case the search space also includes multiple levels of spatial and temporal granularity (\cref{sec:actions}).
The additional complexity is mitigated by an efficient tree search that probabilistically balances exploration and exploitation (\cref{sec:distribution}).

The performance of our algorithm benefits from the heuristic idea to explore simple (coarse granularity) inputs first, then gradually switch to more complex inputs that include finer granularity.
Importantly, the finer granularity tends only to be added where it is needed, thus avoiding the exponential penalty of searching the entire input space at the finer granularity.
While it is always possible to construct pathological problem instances, we find that our approach is very effective on benchmarks from the literature, including the Powertrain benchmark~\cite{Jin-et-al2014}, which is considered a difficult challenge for falsification.
Our experimental results (\cref{sec:experiments}) demonstrate that our algorithm can consistently beat the best existing methods, in terms of speed and reliability of finding a falsifying input.

\section{Preliminaries}
\label{sec:preliminaries}

In this work we represent a deterministic black-box system model as an input/output function~$\M \colon ([0,T] \to \Rn) \to ([0,T] \to \Rm)$.
In general, $\M$ comprises continuous dynamics with discontinuities.
$\M$~takes a time-bounded, real-valued input signal
    $\bu \colon [0,T] \to \Rn$
of length
    $|\bu| = T$
and transforms it to a time bounded output signal~$\by \colon [0,T] \to \Rm$
of the same length, but potentially different dimensionality.
The dimension~$n$ of the input indicates that at each moment
    $t \le T$
within the time horizon~$T$, the value~$\bu(t) \in \Rn$ of the input is an
$n$-dimensional real vector (analogously for the output).

We denote by $\bu_1\bu_2$ the concatenation of signals $\bu_1$ and $\bu_2$ that have the same dimensions, such that
$\bu_1 \cons \bu_2 \colon [0,T_1+T_2] \to \Rn$.
Concatenation of more than two signals follows naturally and is denoted~$\bu_1\bu_2\bu_3\cdots$.
A constant input signal segment is written~$(t,v)$, where~$t$ is a time duration and $v \in \Rn$ is a vector of input values. 
A piecewise constant input signal is the concatenation of such segments.

In this work we adopt the syntax and robustness semantics of STL defined in~\cite{DonzeMaler2010}.
The syntax of an STL formula is thus given by
\begin{equation}
\varphi = \neg\varphi \mid \varphi\lor\varphi \mid \varphi\land\varphi \mid \varphi\Until_I\varphi \mid \square_I\varphi \mid \Diamond_I\varphi \mid \mu, \label{eq:syntax}
\end{equation}
where the logical connectives and temporal operators have their usual Boolean interpretations and equivalences, $I$~is the interval of time over which the temporal operators range,
and atomic formulas $\mu\equiv f(x_1,\dots,x_m)\geq0$ are predicates over the spatial dimensions of a trace.
The robustness of trace~$\by$ with respect to formula~$\varphi$, denoted $\rho(\varphi,\by)$, is calculated inductively according to the following robustness semantics, using the equivalence $\rho(\varphi,\by)\equiv\rho(\varphi,\by,0)$.
\begin{eqnarray*}
\rho(\mu,\by,t)\!\! &{=}&\!\!f(x_1[t],\dots,x_m[t]),\text{for}\,\mu\equiv f(x_1,\dots,x_m)\geq0\\
\rho(\neg\varphi,\by,t) \!\!&{=}&\!\!-\rho(\varphi,\by,t)\\
\rho(\varphi_1{\lor}\varphi_2,\by,t)\!\! &{=}&\!\!\max(\rho(\varphi_1,\by,t),\rho(\varphi_2,\by,t))\\
\rho(\varphi_1{\land}\varphi_2,\by,t) \!\!&{=}&\!\!\min(\rho(\varphi_1,\by,t),\rho(\varphi_2,\by,t))\\
\rho(\varphi_1{\Until_I}\,\varphi_2,\by,t)\!\! &{=}&\!\!\!\! \max_{t'\in t+I}\big(\min_{t''\in[t,t')}(\rho(\varphi_1,\by,t'')), \min(\rho(\varphi_2,\by,t'))\big)\\
\rho(\Diamond_I\varphi,\by,t)\!\! &{=}&\!\!\!\!\max_{t'\in t+I}(\rho(\varphi,\by,t'))\\
\rho(\Box_I\varphi,\by,t)\!\! &{=}&\!\!\!\!\min_{t'\in t+I}(\rho(\varphi,\by,t'))
\end{eqnarray*}

An important characteristic of the robustness semantics is that it is faithful to standard boolean satisfaction, such that
\begin{equation}
\rho(\varphi, \by) > 0 \implies \by \models \varphi
    \quad \text{ and } \quad
\rho(\varphi, \by) < 0 \implies \by \not\models \varphi.
    \label{eq:robustness-sound}
\end{equation}
Together, these equations justify using the robustness semantics $\rho(\varphi,\M(\bu))$ to detect whether an input~$\bu$ corresponds to the violation of a requirement~$\varphi$.
This correspondence is exploited to find such falsifying inputs through global hill-climbing optimization:
\begin{equation}
\text{Find } \bu^* = \argmin_{\bu \in ([0,T] \to \Rn)} \rho(\varphi, \M(\bu)) \text{ such that } \rho(\varphi, \M(\bu^*)) < 0.
    \label{eq:hill-climbing}
\end{equation}
Of course, finding an adequate falsifying input~$\bu^*$ is generally hard and subject to the limitations of the specific optimization algorithm used.

Sound approximations of the lower and upper bounds of the robustness of a prefix~$\by$ can sometimes be used to short-cut the search.
We thus define lower and upper bounds in the following way.
\begin{equation}
\text{Lower:}\;
\ul \rho(\varphi, \by) = \min_{\by'} \rho(\varphi, \by\cons\by')
    \quad\quad
\text{Upper:}\;
\ol \rho(\varphi, \by) = \max_{\by'} \rho(\varphi, \by\cons\by')
    \label{eq:robustness-bounds}
\end{equation}
A lower bound $\ul \rho(\varphi,\M(\bu))) > 0$ can be used to detect that a prefix cannot be extended to a falsifying trace
(e.g., after the deadline for a harmful event has passed).
An upper bound $\ol \rho(\varphi,\M(\bu)) < 0$ similarly implies $\M(\bu\cons\bu') \not\models \varphi$ for all $\bu'$,
concluding that input~$\bu$ is already a witness for falsification (e.g., a limit is already exceeded).

\section{Approach}
\label{sec:approach}

We wish to solve the following falsification problem efficiently:
\begin{equation}
\text{Find } \bu^* \text{ such that } \ol\rho(\varphi, \M(\bu^*)) < 0.
    \label{eq:falsification-problem}
\end{equation}
Our approach is to repeatedly construct input signals
    $\bu = \bu_1 \cons \bu_2 \cons \bu_3 \cons \cdots$,
where~$\bu_i$ is drawn from a predetermined search space of candidate input segments, $\actions$.
The choice is probabilistic, according to a distribution $\distribution$ that determines the search strategy,
i.e., which inputs are likely to be tried next given a partially explored search space.
The construction of each input is done incrementally, to take advantage of the potential short cuts described at the end of~\cref{sec:preliminaries}.

\Cref{alg:prob} ``adaptive Las Vegas Tree Search'' (\aLVTS) codifies the high level functionality of this probabilistic approach, described in detail in \cref{sec:algorithm}.

The effectiveness of our algorithm in practice comes from the particular choices of $\actions$ and~$\distribution$,
which let the search gradually \emph{adapt} to the difficulty of the problem.
The set~$\actions$  (defined in~\cref{sec:actions}) contains input segments of diverse granularity,
which intuitively corresponds to how precise the input must be in order to find a falsifying trace.
The distribution~$\distribution$  (defined in~\cref{sec:distribution}) initially assigns high probabilities to the ``coarsest'' input segments in $\actions$.
Coarse here means that the segments tend to be long in relation to the time horizon~$T$ and large in relation to the extrema of the input space.
The algorithm probabilistically balances exploration and exploitation of segments, but as the coarser segments become fully explored, and the property has not been falsified, the algorithm gradually switches to finer-grained segments.

\subsection{Algorithm}
\label{sec:algorithm}

\begin{algorithm}
    \caption{Adaptive Las Vegas Tree Search (aLVTS)}
    \label{alg:prob}
\SetKw{KwGo}{goto}
\SetKw{KwCont}{continue}
\smallskip
\KwIn{\\
system model~$\M:\bu\rightarrow\by$,\\
\qquad with $\bu:[0,t]\rightarrow\Rn$ and $\by:[0,t]\rightarrow\Rm$\\
time-bounded specification~$\phi$\\
set of all possible input trace segments $\actions$
}
\smallskip
\KwOut{\\
$\bu$ such that $\M(\bu\cons\bu') \not\models \phi$ for all $\bu'$, or\\
$\bot$ after timeout or maximum number of iterations
}
\smallskip
$\unexplored(\bu)\gets\actions$ for all $\bu$
\label{alg:prob:initunexp}

$\explored(\bu)\gets\emptyset$ for all $\bu$
\label{alg:prob:initexp}

\Repeat{timeout or maximum number of iterations}{ \label{alg:prob:repeat}
    $\bu \gets \langle\rangle$
    \label{alg:prob:root}
    \smallskip

    \While{$\unexplored(\bu)\neq\emptyset\vee\explored(\bu)\neq\emptyset$} {
    \label{alg:prob:while}
    
       sample $\bu' \sim \distribution(\bu)$\label{alg:prob:sample}
       \label{alg:prob:sample}
       \smallskip

        \If{$\bu' \in \unexplored(\bu)$} {
        
       $\unexplored(\bu) \gets \unexplored(\bu)\setminus\{\bu'\}$
       \label{alg:prob:remove}

       $\by \gets \M(\bu\cons\bu')$
       \label{alg:prob:output}
       \smallskip

        \If{$\ol\rho(\phi,\by) < 0$} { \label{alg:prob:false}
            \Return {$\bu\cons\bu'$}
            \label{alg:prob:success}
	}
        \smallskip
            \If{$\ul\rho(\phi,\by) > 0$} {\label{alg:prob:true}
           	\KwCont{\cref{alg:prob:repeat}}
           	\label{alg:prob:cont}
           }

         $\explored(\bu) \gets \explored(\bu) \cup \{\bu'\}$
         \label{alg:prob:explored}

        }
        $\bu \gets \bu\cons\bu'$
        \label{alg:prob:extend}
    }
}\label{alg:prob:until}
\Return {$\bot$}

\end{algorithm}

\Cref{alg:prob} searches the space of input signals constructed from piecewise constant (over time) segments,
which are chosen at random according to the distribution defined by $\distribution$ in~\cref{alg:prob:sample}.
This distribution is a function of the numbers of unexplored and explored edges at different levels of granularity, and thus defines the probabilities of exploration, exploitation and adaptation.
The precise calculation made by $\distribution$ is described in~\cref{sec:distribution}.

As the search proceeds, the algorithm constructs a tree whose nodes each correspond to a unique input signal prefix.
The edges of the tree correspond to the constant segments that make up the input signal. 
The root node corresponds to time 0 and the empty input signal (\cref{alg:prob:root}).

To each node identified by an input signal prefix $\bu$ is associated a set of unexplored edges, $\unexplored(\bu) \subseteq \actions$, that correspond to unexplored input signal segments, and a set of explored edges, $\explored(\bu) \subseteq \actions$, that remain inconclusive with respect to falsification.
Initially, all edges are unexplored (\cref{alg:prob:initunexp} and \cref{alg:prob:initexp}).
Once an edge has been chosen (\cref{alg:prob:sample}), the unique signal segment associated to the edge may be appended to the signal prefix associated to the node, to form an extended input signal.
If the chosen edge is unexplored, it is removed from the set of unexplored edges (\cref{alg:prob:remove}) and the extended input signal~$\bu \cons \bu'$ is transformed by the system into an extended output signal (\cref{alg:prob:output}).
If the requirement is falsified by the output signal ($\by$ in \cref{alg:prob:false}), the algorithm immediately quits and returns the falsifying input signal (\cref{alg:prob:success}).
If the requirement is satisfied, with no possibility of it being falsified by further extensions of the signal (\cref{alg:prob:true}), the algorithm quits the current signal (\cref{alg:prob:cont}) and starts a new signal from the root node (\cref{alg:prob:root}).
This is the case, in particular, when the length of the signal exceeds the time horizon of the formula as a consequence of the definition of~$\ul \rho$ in \cref{eq:robustness-bounds}.
If the requirement is neither falsified nor satisfied, the edge is added to the node's set of explored edges (\cref{alg:prob:explored}).
Regardless of whether the chosen edge was previously explored or unexplored, if the signal remains inconclusive, the extended input signal becomes the focus of the next iterative step (\cref{alg:prob:extend}).

\subsection{Definition of $\actions$}
\label{sec:actions}

Given an input signal segment $\bu'$ of length $t$ time units and value $(v_1,\dots,v_n)\in\Rn$, let $\ul v_i$ and $\ol v_i$ denote the minimum and maximum possible values, respectively, of dimension $i\in\{1,\dots,n\}$.
For each integer level $l\in\{0,\dots,l_\mathrm{max}\}$, we define the set of possible proportions of the interval $[\ul v_i,\ol v_i]$ as
$$\bp_l=\{(2j+1)/2^l\mid j\in\N_0\leq(2^l-1)/2\}.$$
The numerators of all elements are coprime with the denominator,~$2^l$, hence $\bp_i\cap  \bp_j=\emptyset$ for all $i\neq j$.
By definition, $\bp_0$ also includes~$0$.
Hence, $\bp_0=\{0,1\}$, $\bp_1 = \left\{{\footnotesize\frac{1}{2}}\right\}$ and $\bp_2 = \left\{{\footnotesize\frac{1}{4}},{\footnotesize\frac{3}{4}}\right\}$, etc.
The set of possible values of dimension $i$ at level $l$ is thus given by
$$\bv_{i,l}=\ul v_i+\bp_l\times(\ol v_i-\ul v_i).$$
Rather than making the granularity of each dimension independent, we interpret the value of $l$ as a granularity ``budget'' that must be distributed among the (non-temporal) dimensions of the input signal.
The set of possible per-dimension budget allocations for level $l$ is given by
$$\bb_l = \{(b_1,...,b_n)\in\N^n_0\mid b_1 + \cdots + b_n = l\}.$$
For example, with $n=2$, $\bb_3=\left\{(0,3),(1,2),(2,1),(3,0)\right\}$.
If we denote the set of possible time durations at level $l$ by $\bt_l$, then the set of possible input segments at level~$l$ is given by
$$
\actions_l=\bigcup_{(b_1,\dots,b_n)\in\bb_l}\bt_l\times\bv_{1,b_1}\times\cdots\times\bv_{n,b_n}.
$$
Note that while $\bt_l$ is not required here to share the granularity budget, this remains a possibility.
Our implementation (\cref{sec:implementation}) actually specifies $\bt_l$ by defining a fixed number of control points per level, $(k_0, \ldots, k_{l_\mathrm{max}})$, such that the $\bt_l = \{T/k_l\}$ are singleton sets.
The sizes of various $\actions_l$ for different choices of $n$ and $l$, assuming $|\bt_l|=1$, are given in \cref{tab:actions}.

\begin{table}
\caption{Size of $\actions_l$ for input dimensionality~$n$ and level~$l$
         given that the size of $\bt_l$ is one.}
\label{tab:actions}
\setlength\tabcolsep{3pt}
\centering
\begin{tabular}{l | rrrrrrrrrrr}
\toprule
$n$     & $l = 0$ & $1$ & $2$ & $3$ & $4$ & $5$ & $6$ & $7$ & $8$ & $9$ & $10$ \\
\midrule
$2$ & 4 & 4 & 9 & 20 & 44 & 96 & 208 & 448 & 960 & 2048 & 4352 \\
$3$ & 8 & 12 & 30 & 73 & 174 & 408 & 944 & 2160 & 4896 & 11008 & 24576 \\
\bottomrule
\end{tabular}
\end{table}

In summary, an input segment $\bu' = (t, v_1, \ldots, v_n)\in\actions_l$ has~$t \in \bt_l$ and corresponding budget allocation $b_1 + \cdots + b_n = l$, with the value~$v_i$ for each dimension given by~$v_i = \ul v_i + p_i (\ol v_i - \ul v_i)$, where $p_i = (2j_i + 1) / 2^{b_i}$, for some~$j_i$, defines the proportion between minimum~$\ul v_i$ and maximum~$\ol v_i$.

By construction, $\actions_i\cap\actions_j=\emptyset$, for all $i\neq j$.
Hence, we define
\begin{eqnarray*}
\unexplored_l(\bu) & = & \unexplored(\bu)\cap\actions_l~\text{and}\\
\explored_l(\bu) & = & \explored(\bu)\cap\actions_l.
\end{eqnarray*}
The set of all possible input signal segments is given by $\actions = \actions_0\cup\actions_1\cup\cdots\cup\actions_{l_\mathrm{max}}$.

\begin{figure*}
\centering
\begin{tikzpicture}[scale=0.87, every fill/.style={black}]
\draw (-0.4,-0.4) rectangle (3.4,3.4);
\fill (0,0) circle (2.5pt);\fill (0,3) circle (2.5pt);\fill (3,0) circle (2.5pt);\fill (3,3) circle (2.5pt);
\draw (0.4,3) node {$\actions_0$};

\draw (3.6,-0.4) rectangle (7.4,3.4);
\fill (4,0) circle (2.5pt);\fill (4,3) circle (2.5pt);\fill (7,0) circle (2.5pt);\fill (7,3) circle (2.5pt);
\fill (5.5,0) circle (2pt);\fill (5.5,3) circle (2pt);\fill (4,1.5) circle (2pt);\fill (7,1.5) circle (2pt);
\draw (4.4,1.5) node {$\actions_1$};

\draw (7.6,-0.4) rectangle (11.4,3.4);
\fill (8,0) circle (2.5pt);\fill (8,3) circle (2.5pt);\fill (11,0) circle (2.5pt);\fill (11,3) circle (2.5pt);
\fill (9.5,0) circle (2pt);\fill (9.5,3) circle (2pt);\fill (8,1.5) circle (2pt);\fill (11,1.5) circle (2pt);
\fill (9.5,1.5) circle (1.5pt);\fill (8,0.75) circle (1.5pt);\fill (8,2.25) circle (1.5pt);\fill (8.75,0) circle (1.5pt);
\fill (8.75,3) circle (1.5pt);\fill (10.25,0) circle (1.5pt);\fill (10.25,3) circle (1.5pt);\fill (11,0.75) circle (1.5pt);
\fill (11,2.25) circle (1.5pt);
\draw (9.9,1.5) node {$\actions_2$};

\draw (11.6,-0.4) rectangle (15.4,3.4);
\fill (12,0) circle (2.5pt);\fill (12,3) circle (2.5pt);\fill (15,0) circle (2.5pt);\fill (15,3) circle (2.5pt);
\fill (13.5,0) circle (2pt);\fill (13.5,3) circle (2pt);\fill (12,1.5) circle (2pt);\fill (15,1.5) circle (2pt);
\fill (13.5,1.5) circle (1.5pt);\fill (12,0.75) circle (1.5pt);\fill (12,2.25) circle (1.5pt);\fill (12.75,0) circle (1.5pt);
\fill (12.75,3) circle (1.5pt);\fill (14.25,0) circle (1.5pt);\fill (14.25,3) circle (1.5pt);\fill (15,0.75) circle (1.5pt);
\fill (15,2.25) circle (1.5pt);
\fill (12.75,1.5) circle (1pt);\fill (13.5,0.75) circle (1pt);\fill (13.5,2.25) circle (1pt);\fill (14.25,1.5) circle (1pt);
\fill (12,0.375) circle (1pt);\fill (12,1.125) circle (1pt);\fill (12,1.875) circle (1pt);\fill (12,2.625) circle (1pt);
\fill (15,0.375) circle (1pt);\fill (15,1.125) circle (1pt);\fill (15,1.875) circle (1pt);\fill (15,2.625) circle (1pt);
\fill (12.375,0) circle (1pt);\fill (12.375,3) circle (1pt);\fill (13.125,0) circle (1pt);\fill (13.125,3) circle (1pt);
\fill (13.875,0) circle (1pt);\fill (13.875,3) circle (1pt);\fill (14.625,0) circle (1pt);\fill (14.625,3) circle (1pt);
\draw (13.9,2.25) node {$\actions_3$};

\draw (15.6,-0.4) rectangle (19.4,3.4);
\fill (16,0) circle (2.5pt);\fill (16,3) circle (2.5pt);\fill (19,0) circle (2.5pt);\fill (19,3) circle (2.5pt);
\fill (17.5,0) circle (2pt);\fill (17.5,3) circle (2pt);\fill (16,1.5) circle (2pt);\fill (19,1.5) circle (2pt);
\fill (17.5,1.5) circle (1.5pt);\fill (16,0.75) circle (1.5pt);\fill (16,2.25) circle (1.5pt);\fill (16.75,0) circle (1.5pt);
\fill (16.75,3) circle (1.5pt);\fill (18.25,0) circle (1.5pt);\fill (18.25,3) circle (1.5pt);\fill (19,0.75) circle (1.5pt);
\fill (19,2.25) circle (1.5pt);
\fill (16.75,1.5) circle (1pt);\fill (17.5,0.75) circle (1pt);\fill (17.5,2.25) circle (1pt);\fill (18.25,1.5) circle (1pt);
\fill (16,0.375) circle (1pt);\fill (16,1.125) circle (1pt);\fill (16,1.875) circle (1pt);\fill (16,2.625) circle (1pt);
\fill (19,0.375) circle (1pt);\fill (19,1.125) circle (1pt);\fill (19,1.875) circle (1pt);\fill (19,2.625) circle (1pt);
\fill (16.375,0) circle (1pt);\fill (16.375,3) circle (1pt);\fill (17.125,0) circle (1pt);\fill (17.125,3) circle (1pt);
\fill (17.875,0) circle (1pt);\fill (17.875,3) circle (1pt);\fill (18.625,0) circle (1pt);\fill (18.625,3) circle (1pt);
\fill(16.75,0.75) circle (0.9pt);\fill(16.75,2.25) circle (0.9pt);\fill(18.25,0.75) circle (0.9pt);\fill(18.25,2.25) circle (0.9pt);
\def\xarray{{16+3/16,16+9/16,16+15/16,16+21/16,16+27/16,16+33/16,16+39/16,16+45/16}}
\foreach \i in {0,...,7}{\pgfmathparse{\xarray[\i]}\fill (\pgfmathresult,3) circle (0.9pt);}
\foreach \i in {0,...,7}{\pgfmathparse{\xarray[\i]}\fill (\pgfmathresult,0) circle (0.9pt);}
\def\yarray{{3/16,9/16,15/16,21/16,27/16,33/16,39/16,45/16}}
\foreach \i in {0,...,7}{\pgfmathparse{\yarray[\i]}\fill (16,\pgfmathresult) circle (0.9pt);}
\foreach \i in {0,...,7}{\pgfmathparse{\yarray[\i]}\fill (19,\pgfmathresult) circle (0.9pt);}
\draw (17.1,2.25) node {$\actions_4$};

\end{tikzpicture}
\caption{Construction of $\actions$ for $n = 2$ and $l_\mathrm{max}=4$, with $\actions_0$ representing the extremes of the two spatial dimensions.
         Showing $\actions_0, \actions_0\cup\actions_1,\dots, \actions_0\cup\actions_1\cup\actions_2\cup\actions_3\cup\actions_4$.
         Larger points correspond to more likely values.}
\label{fig:actions}
\end{figure*}
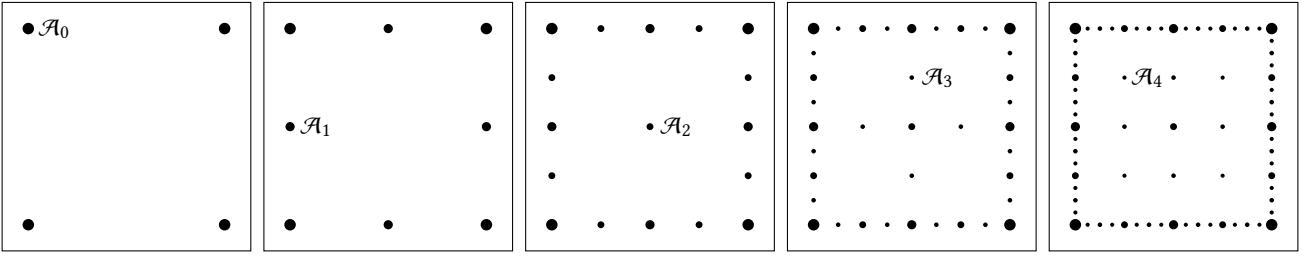

\Cref{fig:actions} depicts the construction of~$\actions$ for two dimensions.
The majority of candidate input points is concentrated on the outer contour,
corresponding to an extreme choice for one dimension and a fine-grained choice for the other dimension.
While this bias appears extreme, as layers are exhausted, finer-grained choices become more likely.
For example, after two points from $\actions_0$ have been tried, all remaining points from both levels in the second panel would be equally probable.

\subsection{Definition of $\distribution$}
\label{sec:distribution}

The distribution $\distribution(\bu)$ is constructed implicitly.
First, a granularity level $l\in\{0,\dots,l_\mathrm{max}\}$ is chosen at random, with probability in proportion to the fraction of edges remaining at the level, multiplied by an exponentially decreasing scaling factor.
Defining the overall weight of level $l$ as
$$w_l=\frac{|\unexplored_l(\bu)|+|\explored_l(\bu)|}{2^l \cdot |\actions_l|},$$ 
level $l$ is chosen with probability $w_l/\sum^{l_\mathrm{max}}_{i=0}w_i$.

Having chosen $l$, one of the following strategies is chosen uniformly at random: 
\begin{enumerate}
\item\label{item:explore} select $\bu'\in\unexplored_l(\bu)$, uniformly at random;
\item select $\bu'\in\explored_l(\bu)$, uniformly at random;
\item\label{item:prefix} select $\bu'\in\explored_l(\bu)$, uniformly at random from those that minimise $\rho(\varphi,\M(\bu\bu'))$;
\item\label{item:suffix} select $\bu'\in\explored_l(\bu)$, uniformly at random from those that minimise $\rho(\varphi,\M(\bu\bu'\bu^*))$, where $\bu^*$ denotes any, arbitrary length input signal suffix that has already been explored from $\bu\bu'$.
\end{enumerate}
Strategy~\ref{item:explore} can be considered pure exploration, while strategies~\ref{item:prefix}--\ref{item:suffix} are three different sorts of exploitation.
In the case that $\unexplored_l(\bu)$ or $\explored_l(\bu)$ are empty, their corresponding strategies are infeasible and a strategy is chosen uniformly at random from the feasible strategies.
If for all $\bu'\in\explored(\bu)$, $\explored(\bu\bu')=\emptyset$, then strategy \ref{item:suffix}  is equivalent to strategy \ref{item:prefix}, but it is {\em not} infeasible.

\subsection{Design Choices}
\label{sec:designchoices}

In contrast to Monte Carlo tree search (MCTS), which typically makes a good decision by reaching a goal state multiple times during playout, our algorithm expects to reach a goal state only once.
In place of a statistic about reaching goal states, we therefore adopt a heuristic, namely the lowest robustness.
Using this heuristic is typically much more effective than choosing entirely at random, but simply substituting the heuristic value for the statistic in the UCT formula (upper confidence bound applied to trees~\cite{KocsisSzepesvari2006}, successfully used with MCTS) is plausible, but nevertheless questionable in our application.
Moreover, the deterministic way that UCT balances exploration and exploitation is often not optimal in the context of falsification.

Our design choices are guided by the observation that simple, coarse-grained input signals either immediately solve typical falsification problems, or need only small modifications to do so.
Since such signals may be quickly explored exhaustively, our level scaling factors are designed to make this happen with high probability.

Unexplored edges have unknown potential, while every explored edge does not yet lead to falsification, but does not exclude it.
The only exception to this occurs when all the edges of a node have been explored and discarded.
In this unlikely case the edge leading to such a node exists and can be traversed, but the node and its trace are immediately rejected (\cref{alg:prob:while}).

While not explicit in the presentation of~\cref{alg:prob}, our approach is deliberately incremental in the evaluation of the system model.
In particular, we can re-use partial simulations to take advantage of the fact that traces share common prefixes.
Hence, for example, one can associate to every visited~$\bu$ the terminal state of the simulation that reached it, using this state to initialize a new simulation when subsequently exploring $\bu\bu'$.
This idea also works for the calculation of robustness.
We note, however, that incremental simulations may be impractical.
For example, suspending and re-starting Simulink can be more expensive than performing an entire simulation from the start.
Under such circumstances, the inner loop of our algorithm can be replaced by a recursion that executes the model on the complete input, returning the trace from the innermost recursion.
The checks in \cref{alg:prob:success,alg:prob:cont} can be done on the appropriate prefix of the output returned from the recursive call (which takes place in \cref{alg:prob:output}).

\section{Evaluation}
\label{sec:experiments}

We briefly describe our implementation and the benchmarks, then present the results of an experimental comparison with random sampling and Breach.

\subsection{Implementation}
\label{sec:implementation}

We have implemented \cref{alg:prob} in the prototype tool~\falstar{},
using the Scala programming language and interfacing to MATLAB/Simulink through a Java API.
\anonymize
    {The code is publicly available on github,%
    \footnote{\url{https://github.com/ERATOMMSD/falstar}}
    including all files and instructions required to re-run our experiments.}
    {The code is publicly available,%
    \footnote{URL omitted for anonymity. A~repeatability package is included in the submission.}
     including all files and instructions required to re-run our experiments.}

The main data structure is an explicit representation of the search tree.
Its nodes are labelled by inputs and the scores relevant for strategies~\ref{item:prefix} and~\ref{item:suffix} of~\cref{sec:distribution}.
Similarly, we keep track of the sets $\unexplored_l$ and $\explored_l$ for each level~$l$.

We do not re-run the simulation for every new input segment, as implied by~\cref{alg:prob:output} of~\cref{alg:prob}. Instead, we cache the output traces within the tree data structure, alongside their inputs.
As mentioned in~\cref{sec:designchoices}, the cost of pausing and resuming a simulation in Simulink incurs a prohibitively large overhead, hence we have implemented a recursive algorithm that executes exactly one simulation up to the full time horizon~$T$ per trial/iteration of the outer loop.

The implementation supports initial parameters with respect to the input signals, such as engine speed~$\omega$ that should remain constant, by augmenting the first level of the search tree with an additional dimension for each of the parameters.

Currently, the possible time durations $\bt_l$ for each level $l$ can be specified statically as singleton sets, using a sequence of numbers of control points $(k_0, \ldots, k_{l_\mathrm{max}})$, such that $\bt_l = \{T/k_l\}$.
The particular choices depend on the requirements of the case study, as noted below.
We make use of both progressive subdivisions (increasing~$k_l$ with~$l$) and constant subdivision (keeping~$k_l$ the same for all~$l$).

We use the algorithm from~\cite{DonzeFerrereMaler2013} to compute the robustness of STL formulas efficiently, and the approach from~\cite{Dreossi2015}
to compute the respective upper and lower bounds.

\falstar{} exposes its functionality through a simple specification and scripting language. In the spirit of the SMT-LIB file format~\cite{smtlib}, the language is S-expression-based, to facilitate integration with other tools. Using this interface one can declare models, input signal specifications, as well as STL requirements.
In addition to solving falsification problems and generating reports, \falstar{} can be used to simulate systems for a given input or to compute robustness values of given STL requirements for specified output traces.
To aid comparisons, \falstar{} can also call Breach as an alternative solver, or generate a standalone MATLAB script to run Breach.

\subsection{Benchmarks}
\label{sec:benchmarks}

The benchmarks here are based on automotive Simulink models that are commonly used in the falsification literature. We focus on those benchmarks that have at least one time-varying input, in contrast to just searching for an initial configuration.

\paragraph{Automatic Transmission}

This benchmark was proposed in~\cite{HoxhaAbbasFainekos2014} and consists of a Simulink model and its related requirements.
The model has two inputs: \emph{throttle} and \emph{brake}. 
Depending on the speed and engine load, an appropriate gear is automatically selected.
Besides the current gear~$g$, the model outputs the speed~$v$ and engine~$\omega$.
We~consider the following requirements:
\begin{align*}
\text{AT1}     &= \Box_{[0,30]}\ v < 120 \\
\text{AT2}~(i) &= \Box_{[0,30]}\ \big(g = i \implies v > 10\cdot i \big) \qquad \text{ for } i \in \{3,4\} \\
\text{AT3}     &= \lnot \big(\Box_{[10,30]} v \in [50,60]\big) \\
\text{AT4}~(\ol v, \ul \omega) &= \big(\Box_{[0,10]}\ v < \ol v\big) \lor \big(\Diamond_{[0,30]}\ \ul \omega < \omega\big)
\end{align*}
There are a few noteworthy details.
The robustness score of~AT2 remains~$\infty$ until $g = i$ is found by unguided search,
and will jump abruptly to a finite value at that time.
To falsify~AT3, a fairly precise speed between 50 and 60 has to be reached and maintained,
which requires a fine grained control of the input, which could be a challenge for \aLVTS.
To falsify~AT4 one must balance keeping the engine~RPM $\omega \le \ul \omega$ while accelerating sufficiently to reach speed $v > \ol v$ before the deadline at time~$10$.

The input signal for the benchmarks is piecewise constant with 4~control points for random sampling/Breach,
which is sufficient to falsify all requirements.
We choose 6 levels, with 2,2,3,3,3,4 control points, respectively,
corresponding to a time granularity of input segment durations between~15 (= $\frac{30}{2}$, coarsest) to 7.5 (= $\frac{30}{4}$, finest).

\paragraph{Powertrain Control}

The benchmark was proposed for hybrid systems verification and falsification in~\cite{Jin-et-al2014}.
It comprises a detailed model of a fuel control system that maintains the air-to-fuel $\AF$ ratio close to a reference value $\AFref$.
The model has two control algorithms:
During startup and when high power is requested, it operates in feed-forward mode,
whereas otherwise measurements of the actual air-to-fuel are fed back into the controller.%
Falsification tries to detect amplitude and duration of spikes in the air-to-fuel ratio that occur as a response either to mode switches or to changes in the throttle~$\theta$ at a given engine speed~$\omega$.
The relation between the three quantities is nontrivial and in part determined by time delays.
The input~$\theta \in [0,62.1)$ (normal mode) and~$\theta \in [61.2,81.2)$ (power mode) varies throughout the trace,
whereas $\omega \in [900,1100]$ is initially chosen and kept constant.
Requirements are expressed with respect to the normalized error $\mu = |\AF - \AFref| / \AFref$,
where $\AFref$ depends dynamically on the current mode.

We consider requirement 27 from~\cite{Jin-et-al2014}, denoted AFC27 in the following, which states that after a falling or rising edge of the throttle, $\mu$ should return to the reference value within~$1$ time unit and stay close to it for some time. We represent this as
\begin{align*}
\text{AFC27}         &= \Box_{[11,50]}\ (\mathit{rise} \lor \mathit{fall}) \implies \Box_{[1,5]}\ |\mu| < 0.008,
\end{align*}
where rising and falling edges of~$\theta$ are detected by
$\mathit{rise} = \theta <  8.8 \land \Diamond_{[0,\epsilon]}\ 40.0 < \theta$ and
$\mathit{fall} = 40.0 < \theta \land \Diamond_{[0,\epsilon]}\ \theta <  8.8$
for $\epsilon = 0.1$.
The concrete bound of 0.008 in~AFC27 and the edge detection parameters are taken from the report of the ARCH friendly competition on falsification
(which featured this problem in~2017~\cite{Dokhanchi-et-al2017} and 2018~\cite{Dokhanchi-et-al2018}) as a balance between difficulty of the problem and ability to find falsifying traces.
The interval $[11,50]$ over which the requirement is checked begins~$1$ time unit after the transition from startup to normal mode at time~$10$.
Without this short delay, falsification would trivially discover the large spike following the mode switch. 

The input signal is piecewise constant with 10~control points, again following~\cite{Dokhanchi-et-al2017,Dokhanchi-et-al2018}, and 5~levels with 10~control points each for \aLVTS.

\subsection{Experimental Results}
\label{sec:results}

We compare the performance of \aLVTS\ with uniform random sampling (both implemented in \falstar
\anonymize{\footnote{\url{https://github.com/ERATOMMSD/falstar} commit 43f5ca7}}{})
and with with the state-of-the-art stochastic global optimization algorithm CMA-ES~\cite{Igel2007} implemented in the falsification tool Breach.%
    \footnote{\url{https://github.com/decyphir/breach} release version~1.2.9}
We do not make a comparison with Breach's Nelder-Mead algorithm, since it has significantly poorer performance than CMA-ES on almost all of the benchmarks.
The machine and software configuration was:
CPU Intel~i7-3770, 3.40GHz, 8~cores, 8Gb~RAM,
64-bit Ubuntu~16.04 kernel~4.4.0,
MATLAB~R2018a, Scala~2.12.6, Java~1.8.

We compare two performance metrics:
\emph{success rate} (how many falsification trials were successful in finding a falsifying input)
and the \emph{number of iterations} made, which corresponds to the number of simulations required and thus indicates time.
To account for the stochastic nature of the algorithms, the experiments were repeated for $50$~trials.
For a meaningful comparison of the number of iterations until falsification,
we tried to maximize the falsification rate for a limit of $300$~iterations per trial.

The number of iterations of the top-level loop in \cref{alg:prob} in our implementation corresponds exactly to one complete Simulink simulation up to the time horizon~(cf.~\cref{sec:implementation}).
For random sampling and CMA-ES, the number of iterations likewise corresponds to samples taken by running exactly one simulations each.
Hence the comparison is fair and, as the overhead is dominated by simulation time, the numbers are roughly proportional to wall-clock times.

\newcommand{\CB}{\cellcolor{blue!20}}
\newcommand{\CG}{\cellcolor{gray!40}}

\begin{table}[t]
\setlength\tabcolsep{2.5pt}
    \caption{Experimental Results comparing the
             success rate out of of 50 trials (``succ.'' , higher is better) and
             number of iterations averaged over successful trials (``iter.'', lower is better, where M=arithmetic mean, SD=standard deviation)
             of uniform random sampling, Breach/CMA-ES, and \falstar/\cref{alg:prob}
             for a maximum of 300~iterations per trial.
             The best results for each requirement are highlighted.
    }
    \label{tab:results}
\begin{tabular}{lrrrrrrrrrr}
\toprule
	& \multicolumn{3}{c}{\bf}
	& \multicolumn{3}{c}{\bf Breach:}
	& \multicolumn{3}{c}{\bf \falstar:} \\
	& \multicolumn{3}{c}{\bf Random}
	& \multicolumn{3}{c}{\bf CMA-ES}
	& \multicolumn{3}{c}{\bf \aLVTS} \\
	      & \multicolumn{1}{c}{succ.} & \multicolumn{2}{c}{iter.}
	      & \multicolumn{1}{c}{succ.} & \multicolumn{2}{c}{iter.}
          & \multicolumn{1}{c}{succ.} & \multicolumn{2}{c}{iter.} \\
Formula   & \multicolumn{1}{c}{/50} & \multicolumn{1}{c}{M} & \multicolumn{1}{c}{SD}
          & \multicolumn{1}{c}{/50} & \multicolumn{1}{c}{M} & \multicolumn{1}{c}{SD}
          & \multicolumn{1}{c}{/50} & \multicolumn{1}{c}{M} & \multicolumn{1}{c}{SD} \\
      \cmidrule(r){1-1}
      \cmidrule(rl){2-4}
      \cmidrule(rl){5-7}
      \cmidrule(l){8-10}
AT1         &    43 &  106.6 & 83.9 &\CB 50 &    39.7 &    23.6 &\CB  50 &\CB  8.5  &\CB  6.7 \\      
AT2~$(i=3)$ &\CB 50 &   41.0 & 36.7 &\CB 50 &\CB 13.2 &\CB  9.1 &\CB  50 &    33.4  &    27.5 \\      
AT2~$(i=4)$ &    49 &   67.0 & 60.8 &     6 &    17.8 &    15.9 &\CB  50 &\CG 23.4  &\CG 22.5 \\      
AT3         &    19 &  151.1 & 98.1 &\CB 50 &   145.2 &    63.0 &\CB  50 &\CB 86.3  &\CB 52.1 \\      
AT4~$(a)$   &    36 &  117.3 & 71.8 &\CB 50 &    97.0 &    47.7 &\CB  50 &\CB 22.8  &\CB 10.6 \\      
AT4~$(b)$   &     2 &  117.7 &  9.2 &    49 &    46.7 &    58.0 &\CB  50 &\CG 47.6  &\CB 23.5 \\      
      \cmidrule(r){1-1}
      \cmidrule(l){2-4}
      \cmidrule(l){5-7}
      \cmidrule(l){8-10}
Summary AT  &   199 &   95.3 & 47.9 &   255 &    42.8 &    29.0 &\CB 300 &\CB 29.2  &\CB 19.4 \\
\midrule
AFC27       &    15 &  129.1 & 90.8 &    41 &  121.0 & 49.3 &\CB  50 &\CB  3.9      &\CB  4.3 \\      
\bottomrule
\end{tabular}
Parameters
AT4~$(a)$: $\ol v = 80, \ul \omega =4500$, 
AT4~$(b)$: $\ol v = 50, \ul \omega =2700$.
Summary AT
for succ.: sum,
for iter.: geometric mean (see \cite{FlemingW1986}).
\end{table}

\Cref{tab:results} summarizes our results in terms of success rate, and average number of iterations (M) and standard deviation (SD) of successful trials.
The unambiguously (possibly equal) best results are highlighted in blue.
Where the lowest average number of iterations was achieved without finding a falsifying input for every trial, we highlight in grey the lowest average number of iterations for $100\%$ success.
We thus observe that \aLVTS\ achieves the best performance in all but one case, AT2~$(i=3)$.
Importantly, within the budget of~300 iterations per trial, \aLVTS\ achieves a perfect success rate.
CMA-ES is successful in 296~trials out of the total~350, with sub maximal success for AT2~$(i=4)$ and AFC27.
In comparison, random sampling succeeds in only~214 trials, with sub maximal success in all but AT2.

The number of iterations required for falsification varies significantly between the algorithms and between the benchmarks.
For the automatic transmission benchmarks, as an approximate indication of relative performance,
CMA-ES requires about 50\%~more iterations as \aLVTS\ (geometric mean: 42.8 vs. 29.2), and random sampling requires again twice as many as CMA-ES.
Except for AT2, variance of \aLVTS\ is lower than that of CMA-ES.

Looking at the individual benchmarks, CMA-ES is only consistently faster than \aLVTS\ on AT2~$(i=3)$.
Despite having fewer average iterations on AT2~$(i=4)$, CMA-ES cannot be considered faster than \aLVTS\ because most of its trials fail to find a falsifying input---the numbers of iterations are not comparable.
For the powertrain model (AFC27), the performance of \aLVTS\ is more than an order of magnitude better: 3.9~iterations on average, compared to 121 for CMA-ES.

\begin{figure}
\includegraphics[width=0.48\textwidth]{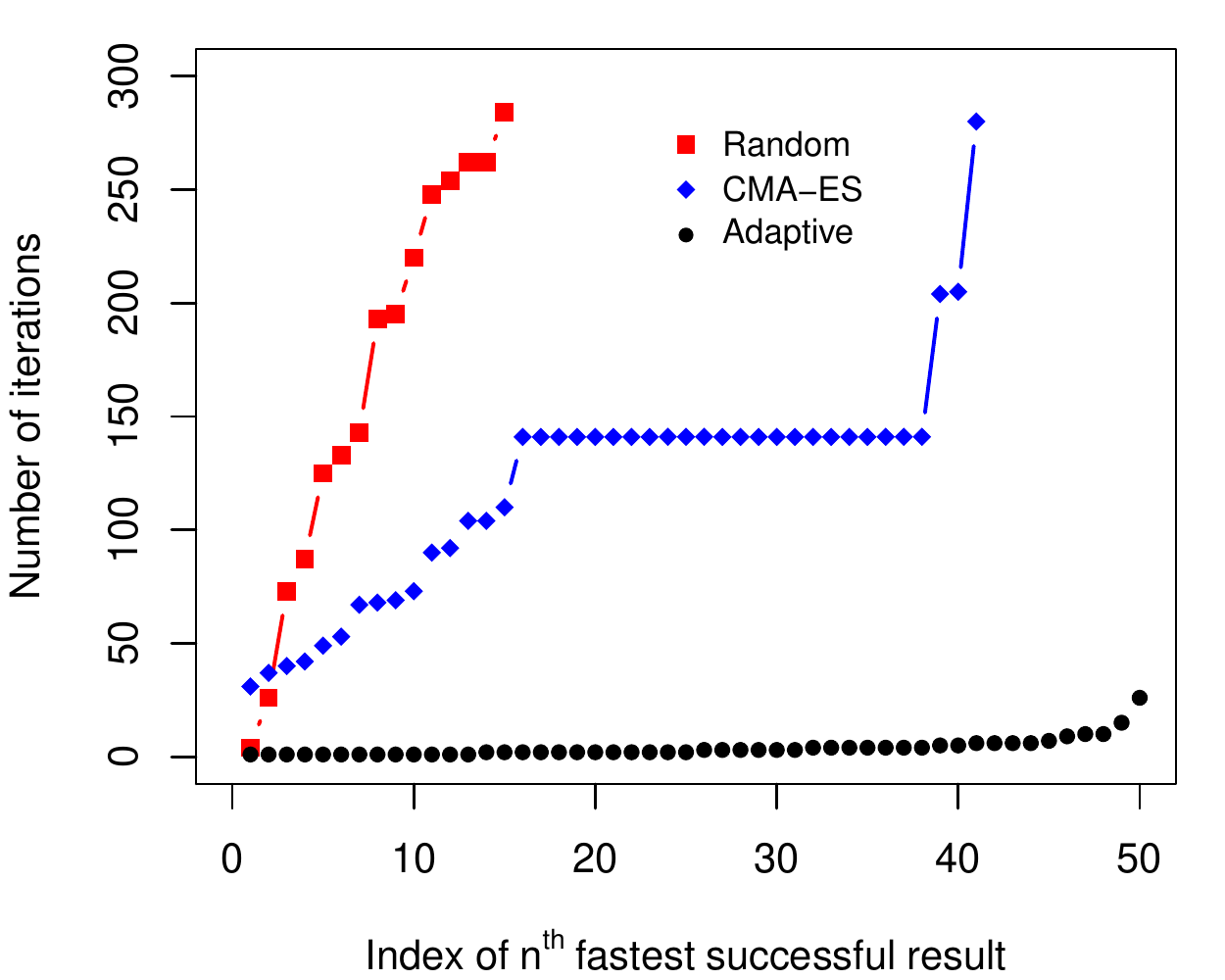}
\caption{Relative performance of falsification algorithms with AFC27 powertrain requirement: fewer iterations and more successful results are better.}
\label{fig:afc27}
\end{figure}

\Cref{fig:afc27} compares all trial runs for AFC27, ordered by the number of iterations required for falsification.
Similar plots for the automatic transmission benchmarks are shown in~\cref{fig:at}.
The shape of each curve gives an intuition of the performance and consistency of its corresponding algorithm.
In general, fewer iterations and more successful results are better, so it is immediately clear that our \aLVTS\ performs significantly better than both random sampling and CMA-ES.

\subsection{Discussion}
\label{sec:discussion}

\begin{figure*}[h]
\caption{Performance comparison for the automatic transmission benchmark}
\label{fig:at}
\centering
    \begin{subfigure}[b]{0.45\textwidth}
    \includegraphics[width=\textwidth]{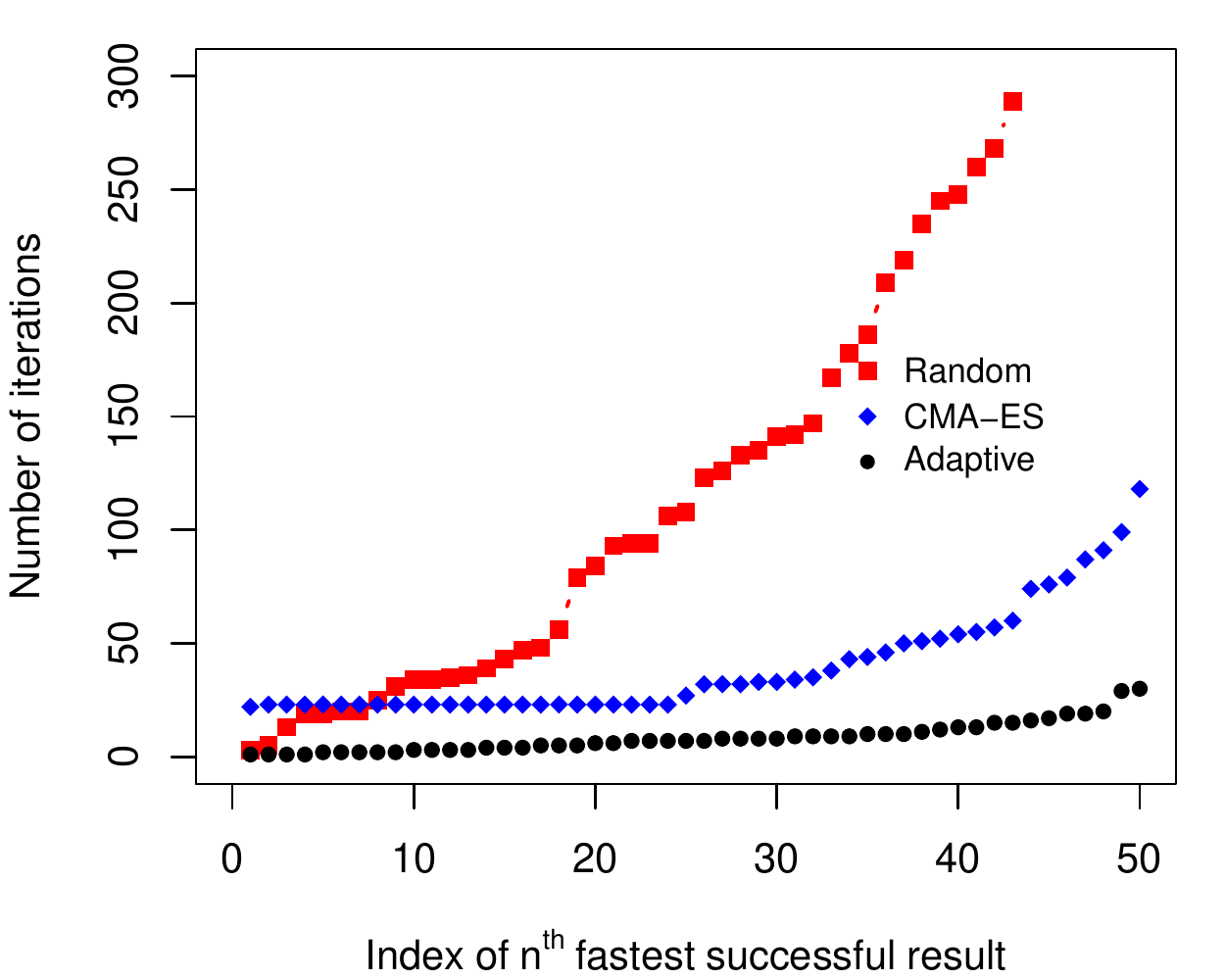}
    \caption{Performance plot for AT1}
    \label{fig:at1}
    \end{subfigure}
\qquad
    \begin{subfigure}[b]{0.45\textwidth}
    \includegraphics[width=\textwidth]{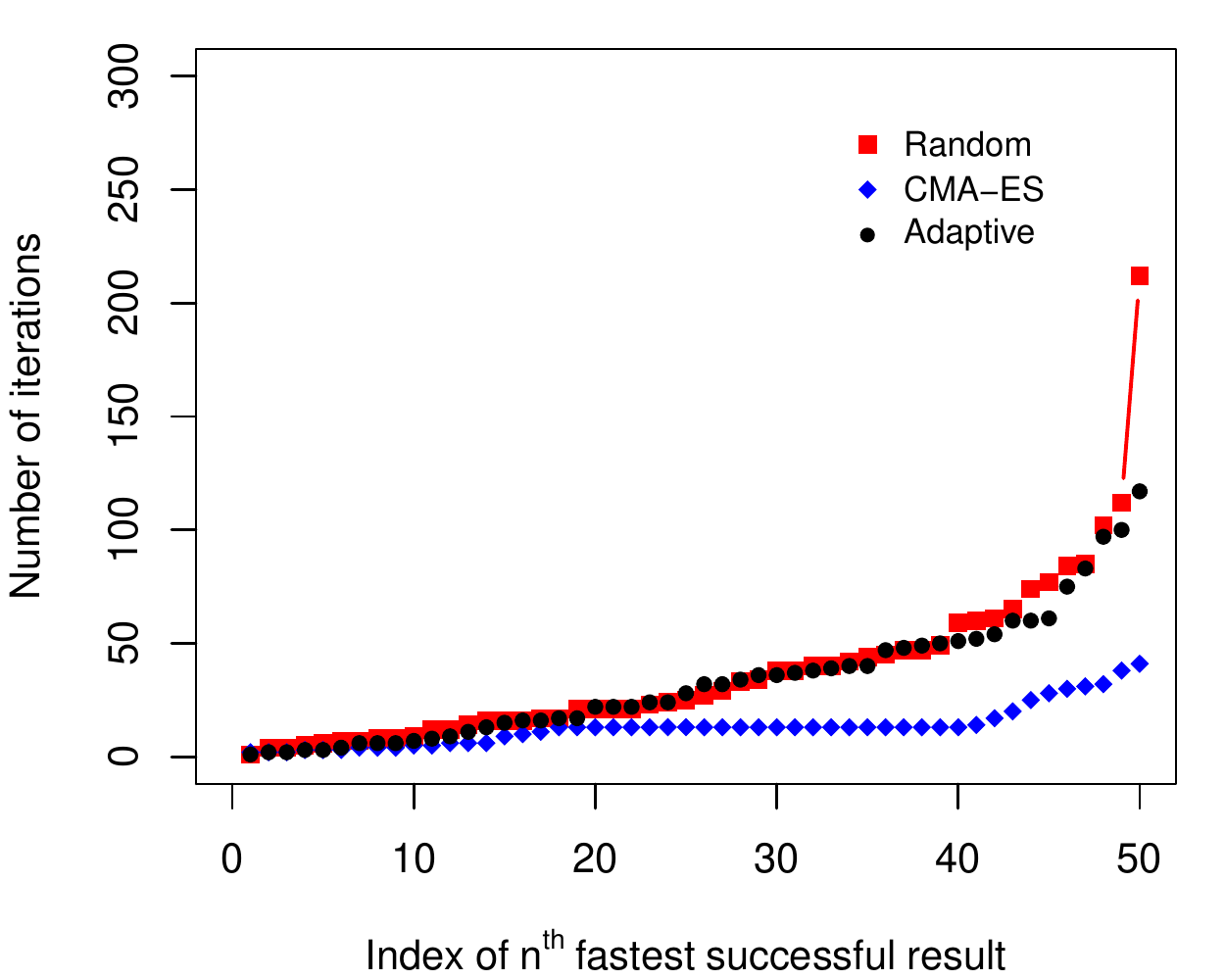}
    \caption{Performance plot for AT2 $(i = 3)$}
    \label{fig:at2-easy}
    \end{subfigure}
\qquad
    \begin{subfigure}[b]{0.45\textwidth}
    \includegraphics[width=\textwidth]{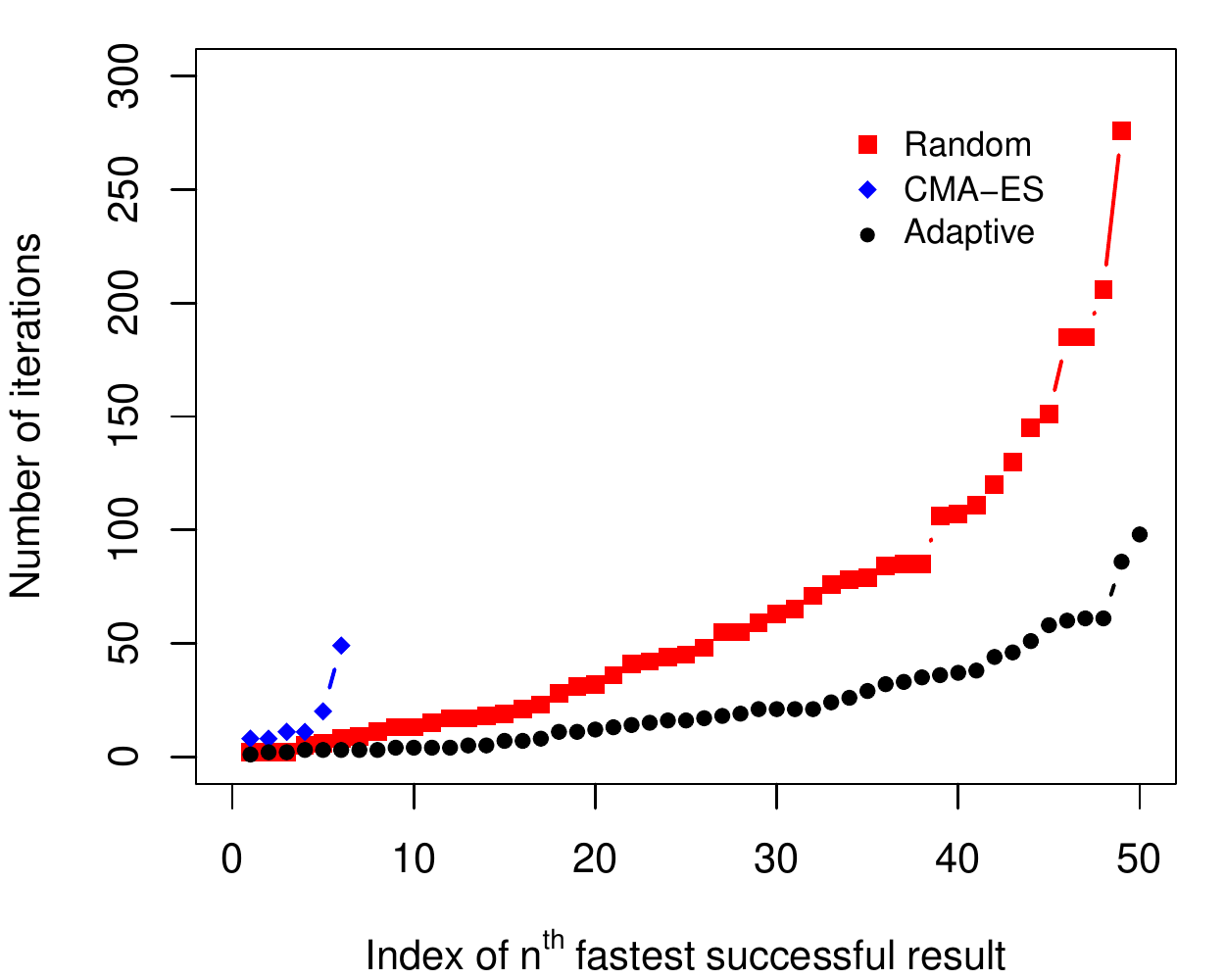}
    \caption{Performance plot for AT2 $(i = 4)$}
    \label{fig:at2-hard}
    \end{subfigure}
\qquad
    \begin{subfigure}[b]{0.45\textwidth}
    \includegraphics[width=\textwidth]{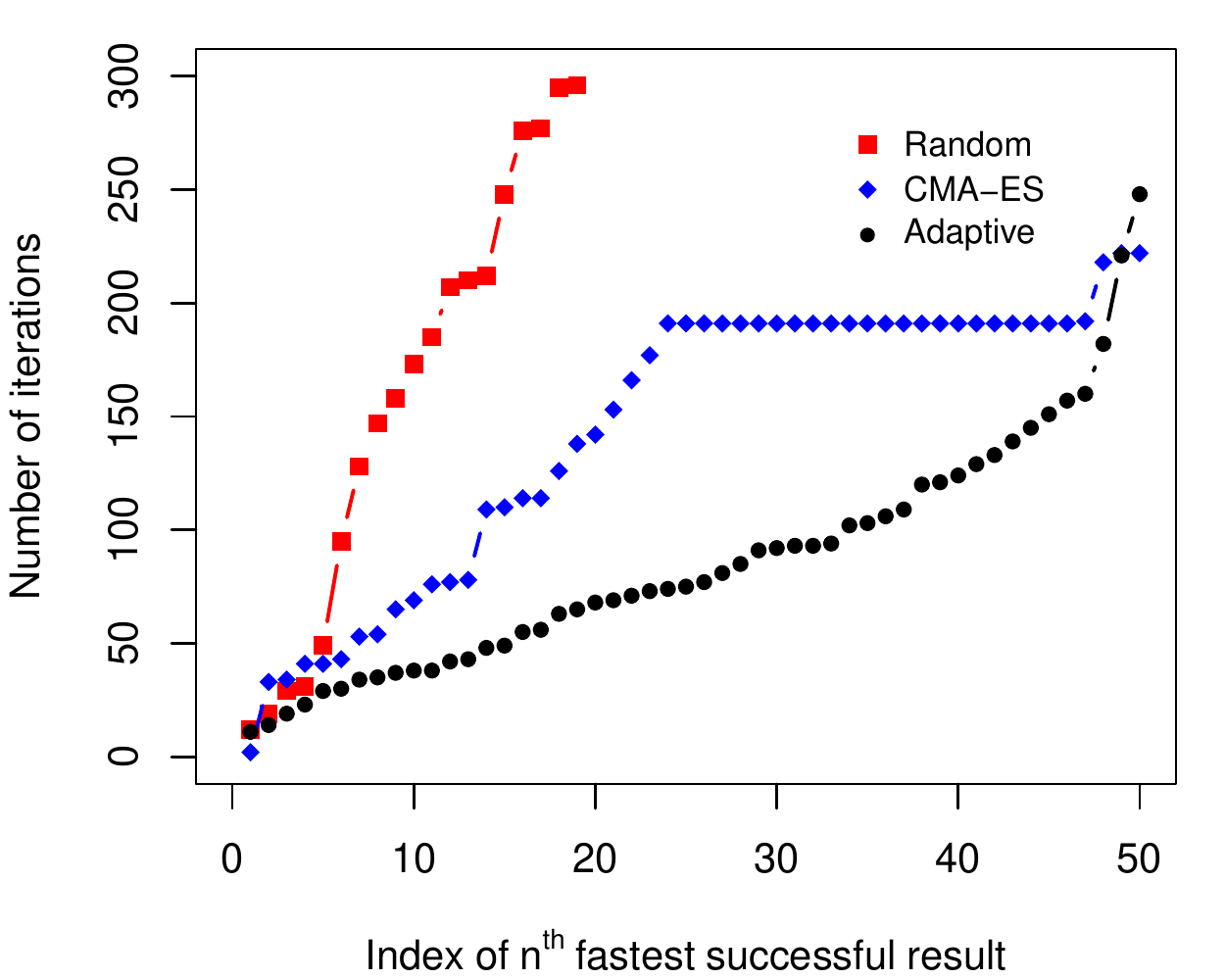}
    \caption{Performance plot for AT3}
    \label{fig:at3}
    \end{subfigure}
\qquad
    \begin{subfigure}[b]{0.45\textwidth}
    \includegraphics[width=\textwidth]{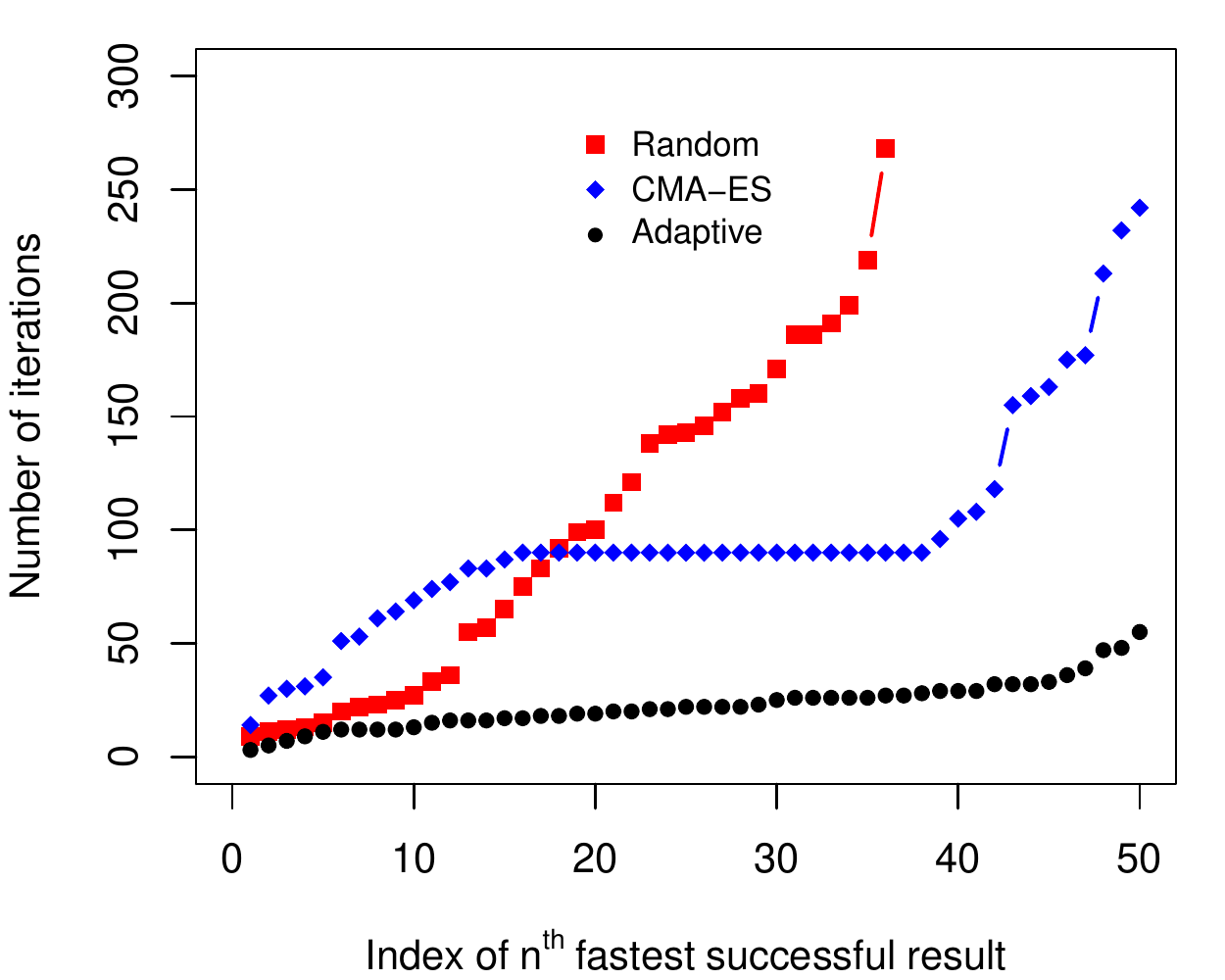}
    \caption{Performance plot for AT4~$(\ol v = 80, \ul \omega = 4500)$}
    \label{fig:at4-easy}
    \end{subfigure}
\qquad
    \begin{subfigure}[b]{0.45\textwidth}
    \includegraphics[width=\textwidth]{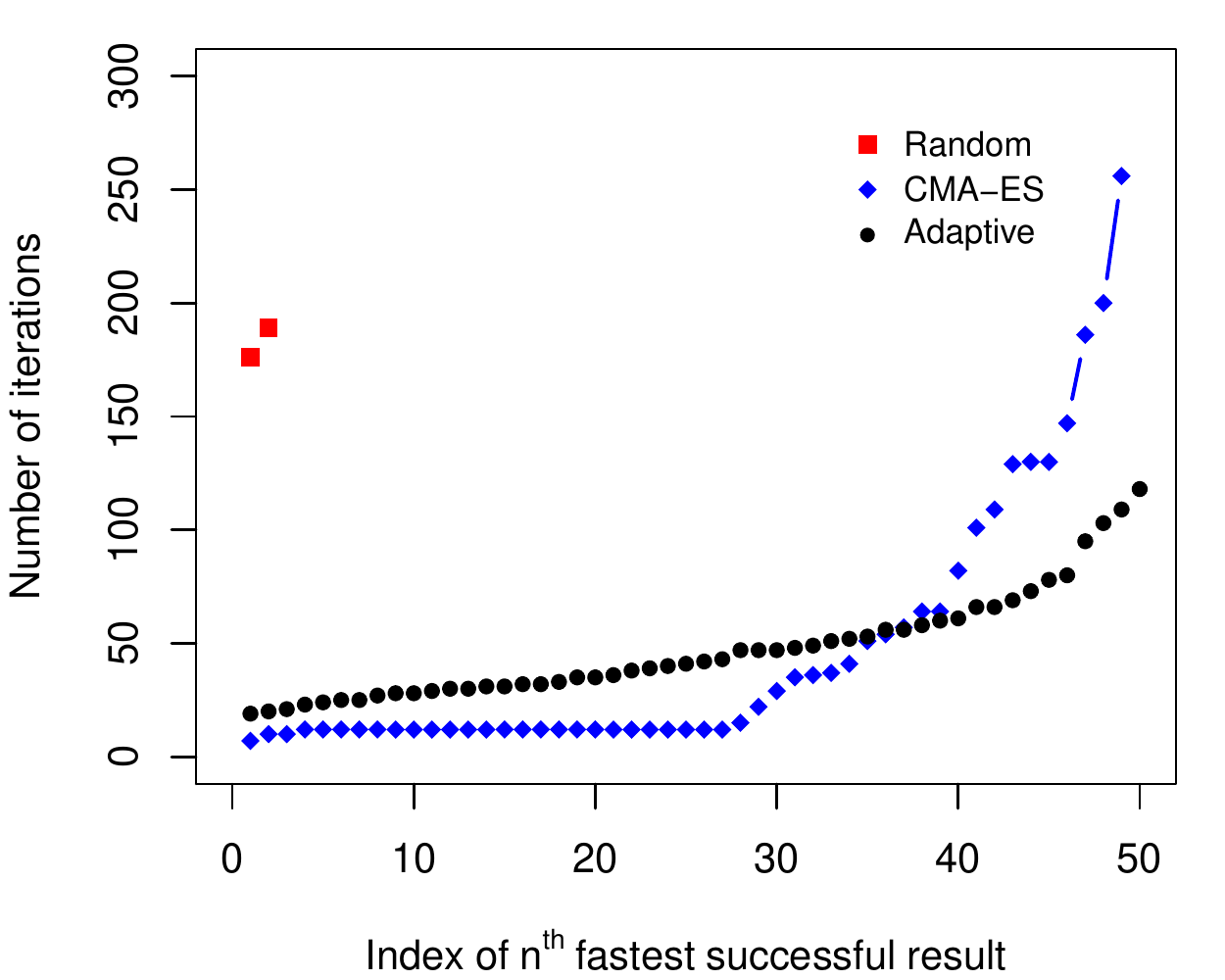}
    \caption{Performance plot for AT4~$(\ol v = 50, \ul \omega = 2700)$}
    \label{fig:at4-hard}
    \end{subfigure}
\end{figure*}

For AT1, \aLVTS\ quickly finds the falsifying input signal, as the required throttle of~$100$ and brake of~$0$ are contained in level~0 and are very likely to be tried early on.
In contrast, even though this is a problem that is well-suited to hill-climbing, CMA-ES has some overhead to sample its initial population, as clearly visible in \cref{fig:at1}.

While CMA-ES deals very well with AT2 for $i=3$, it struggles to find falsifying inputs for $i=4$ (cf. \cref{fig:at2-easy,fig:at2-hard}).
We attribute this to the fact that reaching gear 4 by chance occurs rarely in the exploration of CMA-ES when the robustness score is uninformative.
\aLVTS\ not only explores the spatial dimensions, but takes opportunistic jumps to later time points, which increases the probability of discovering a trace (prefix) where the gear is reached.

A priori, one would expect CMA-ES to perform well with AT3 and AT4, exploiting its continuous optimization to fine tune inputs between conflicting requirements.
E.g., AT3 requires that $v$ is both above~$50$ and below~$60$; AT4 requires that $v$ is high while maintaining low $\omega$, which is proportional to $v$.
One would similarly expect the limited discrete choices made by \aLVTS\ to hinder its ability to find falsifying inputs.
Despite these expectations, our results demonstrate that in most situations \aLVTS\ converges to a falsifying input more consistently and with fewer iterations than CMA-ES.
We speculate that this is because CMA-ES is too slow to reach the ``sweet spots'' in the input space, where its optimization is efficient.

For~AT3, there are a few instances where where \aLVTS\ does not quickly find a good prefix towards the corridor $v \in [50,60]$ at time~10 (the rightmost points in \cref{fig:at3}),
which can be explained by the probabilistic nature of the search.

Regarding two instances of~AT4 in~\cref{fig:at4-easy,fig:at4-hard}, , the graph for \aLVTS\ is generally smoother and shallower,
whereas CMA-ES shows consistent performance for only some of the trials but takes significantly more time on the worst 10 trials.
We remark that the two parameter settings seem to pose opposite difficulty for the two algorithms, as CMA-ES is significantly quicker for two thirds of the trials for the second instance.
It is unclear what causes this variation in performance of CMA-ES.

The plateaux apparent in some of the results using CMA-ES are difficult to explain, but suggest some kind of procedural logic or counter to decide termination.
In contrast, the curves for random sampling and \aLVTS\ are relatively smooth, reflecting their purely probabilistic natures.

\section{Related Work}
\label{sec:related}

The idea to find falsifying inputs using robustness as an optimization function for global optimization originates from~\cite{FainekosPappas2009journal}
and has since been extended to the parameter synthesis problem (e.g.,~\cite{JinDDS2015}).
Two mature implementations in MATLAB are S-Taliro~\cite{Annpureddy-et-al2011} and Breach~\cite{Donze2010},
which have come to define the benchmark in this field.
We describe their optimization algorithms below.
Approaches to make the robustness semantics more informative are~\cite{EddelandMFA2017,AkazakiHasuo2015}.
These use integrals instead of min/max in the semantics of temporal operators.

Underminer~\cite{Underminer} is a recent falsification tool that learns the
(non-) convergence of a system to direct falsification and parameter mining.
It supports STL formulas, SVMs, neural nets, and Lyapunov-like functions as
classifiers.
Other global approaches include~\cite{AdimoolamDDKJ17}, which partitions the input space into sub-regions from which falsification trials are run selectively. This method uses coverage metrics to balance exploration and exploitation.
Comprehensive surveys of simulation based methods for the analysis of hybrid systems are given in~\cite{KDJIB16,BDDFMNS2018}.

Users of S-Taliro and Breach can select from a range of optimization algorithms (Uniform Random, Nelder-Mead, Simulated Annealing, Cross-Entropy, CMA-ES).
These cover a variety of trade-offs between exploration of the search space and exploitation of known good intermediate results.
As such, their performance varies with the structure of the problem at hand.

In general, global optimization needs to find a solution in an \emph{unstructured} combinatorial search space, where the different parameters are assumed, a priori, to be independent.
Good combinations of parameter choices have to be discovered first, before the optimization can work effectively (cf. AT2 and AT3).
This incurs an exponential cost of exhausting the different combinations, which is mitigated in different ways by the various algorithms.
Nelder-Mead is seeded by a constant number of samples and tries to obtain better points in the search space by making linear combinations of the previous ones, according to a fixed scheme. The price paid is not just lack of exploration, but also the danger of selecting mid points between good and {\em bad} values in some dimensions, as a consequence of the weighted average of linear combinations.
To this end, Cross-Entropy methods and CMA-ES discover relations between parameters and exploit these by decorrelating the choices across independent dimensions (i.e., not trying all possible combinations),
as explained nicely in the context of falsification in~\cite{SankaranarayananFainekos2012}.
Simulated Annealing, on the other hand, relies on fairly unguided random exploration, and thus requires many iterations to converge, which can be prohibitively expensive for complex models.

In comparison, our approach explicitly takes into account the temporal structure of the falsification problem.
The exponential search space is pruned by focusing on the early part of the input signal, by investigating a small number of coarse choices first, and by ignoring those prefixes which are unpromising according to the robustness heuristic.
As demonstrated by our experiments, the strength of optimization algorithms to discover precise solutions in the continuous input space does not seem efficient when the mass of falsifying traces is large and exploration is the key factor.

The characteristic of our approach to explore the search space incrementally is shared with {\em rapidly-exploring random trees} (RRTs).
The so-called star discrepancy metric guides the search towards unexplored regions and a local planner extends the tree at an existing node with a trajectory segment that closely reaches the target point.
RRTs~have been used successfully in robotics~\cite{LaValleKuffner2001} and also in falsification~\cite{Dreossi2015}.
On the other hand, the characteristic of taking opportunistic coarse jumps in time is reminiscent of stochastic local search~\cite{DeshmukhJKM2015} and multiple-shooting~\cite{ZutshiDSK14}.

Monte Carlo tree search (MCTS) has been applied to a model of aircraft collisions in~\cite{LKMBO15};
and in a falsification context more recently~\cite{ZESAH18} to guide global optimization,
building on the previous idea of time-staging~\cite{ZEHS18}.
That work noted the strong similarities between falsification using MCTS and statistical model checking (SMC) using \emph{importance splitting}~\cite{JegourelLegaySedwards2013}.
The robustness semantics of STL, used in~\cite{ZEHS18,ZESAH18} and the present approach to guide exploration, can be seen as a ``heuristic score function''~\cite{JegourelLegaySedwards2014} in the context of importance splitting.
All these approaches construct trees from traces that share common prefixes deemed good according to some heuristic.
The principal difference is that importance splitting aims to construct a diverse set of randomly-generated traces that all satisfy a property (equivalently, falsify a negated property), while falsification seeks a single falsifying input.
The current work can be distinguished from standard MCTS and reinforcement learning~\cite{SB2018} for similar reasons. These techniques tend to seek optimal policies that make good decisions in \emph{all} situations, unnecessarily covering the entire search space.

\section{Conclusion}
\label{sec:conclusion}

We have presented a probabilistic algorithm that finds inputs to a hybrid system that falsify a given temporal logic requirement.
On standard benchmarks our algorithm significantly and consistently outperforms existing methods.
While the falsification problem is inherently hard (no theoretically best solution can exist),
we believe that we have found an approach that gives good results in practice, with probabilistic certainty.
We believe the reason it works well stems from a property shared by many falsification problems, namely, that there tends to be a significant mass of falsifying inputs.
By design it scales with the difficulty of the problem, finding trivial solutions (extreme inputs) immediately and has a small constant overhead for ``almost easy'' problems.
This pays off, in particular, for system models that are expensive to simulate, such as the powertrain benchmark. The approach admits incremental computation of simulations and can easily be parallelized.

As future work, we wish to extend the method with some form of optimization, e.g., by interpolation of previously seen traces, reminiscent of the linear combinations computed by Nelder-Mead. We expect that this would help to ``fine-tune'' results with small positive robustness into full solutions, mitigating the limitation inherent in the discretization of the input space.
Likewise, extrapolation of results could be used to propagate the robustness from one level to the next, reducing the need to look at different branches with similar inputs.
Finally, we plan to apply the approach to more benchmarks, specifically those that combine discrete and continuous input domains.

\begin{acks}
The authors are supported by ERATO HASUO Metamathematics for Systems Design Project
(No.~\grantnum{http://dx.doi.org/10.13039/501100009024}{JPMJER1603}),
\grantsponsor{JST}{Japan Science and Technology Agency}{http://www.jst.go.jp};
and Grants-in-Aid \grantnum{JSPS}{No.~15KT0012},
\grantsponsor{JSPS}{Japan Society for the Promotion of Science}{https://www.jsps.go.jp}.
\end{acks}

\bibliographystyle{ACM-Reference-Format}
\bibliography{main}


\begin{thebibliography}{33}


\ifx \showCODEN    \undefined \def \showCODEN     #1{\unskip}     \fi
\ifx \showDOI      \undefined \def \showDOI       #1{#1}\fi
\ifx \showISBNx    \undefined \def \showISBNx     #1{\unskip}     \fi
\ifx \showISBNxiii \undefined \def \showISBNxiii  #1{\unskip}     \fi
\ifx \showISSN     \undefined \def \showISSN      #1{\unskip}     \fi
\ifx \showLCCN     \undefined \def \showLCCN      #1{\unskip}     \fi
\ifx \shownote     \undefined \def \shownote      #1{#1}          \fi
\ifx \showarticletitle \undefined \def \showarticletitle #1{#1}   \fi
\ifx \showURL      \undefined \def \showURL       {\relax}        \fi
\providecommand\bibfield[2]{#2}
\providecommand\bibinfo[2]{#2}
\providecommand\natexlab[1]{#1}
\providecommand\showeprint[2][]{arXiv:#2}

\bibitem[\protect\citeauthoryear{Adimoolam, Dang, Donz{\'e}, Kapinski, and
  Jin}{Adimoolam et~al\mbox{.}}{2017}]%
        {AdimoolamDDKJ17}
\bibfield{author}{\bibinfo{person}{Arvind Adimoolam}, \bibinfo{person}{Thao
  Dang}, \bibinfo{person}{Alexandre Donz{\'e}}, \bibinfo{person}{James
  Kapinski}, {and} \bibinfo{person}{Xiaoqing Jin}.}
  \bibinfo{year}{2017}\natexlab{}.
\newblock \showarticletitle{Classification and Coverage-Based Falsification for
  Embedded Control Systems}. In \bibinfo{booktitle}{\emph{Computer Aided
  Verification}}, \bibfield{editor}{\bibinfo{person}{Rupak Majumdar} {and}
  \bibinfo{person}{Viktor Kun{\v{c}}ak}} (Eds.). \bibinfo{publisher}{Springer},
  \bibinfo{address}{Cham}, \bibinfo{pages}{483--503}.
\newblock


\bibitem[\protect\citeauthoryear{Akazaki}{Akazaki}{2016}]%
        {Akazaki2016}
\bibfield{author}{\bibinfo{person}{Takumi Akazaki}.}
  \bibinfo{year}{2016}\natexlab{}.
\newblock \showarticletitle{Falsification of Conditional Safety Properties for
  Cyber-Physical Systems with Gaussian Process Regression}.
\newblock In \bibinfo{booktitle}{\emph{Runtime Verification (RV)}},
  \bibfield{editor}{\bibinfo{person}{Yli{\`{e}}s Falcone} {and}
  \bibinfo{person}{C{\'{e}}sar S{\'{a}}nchez}} (Eds.). \bibinfo{series}{LNCS},
  Vol.~\bibinfo{volume}{10012}. \bibinfo{publisher}{Springer},
  \bibinfo{pages}{439--446}.
\newblock


\bibitem[\protect\citeauthoryear{Akazaki and Hasuo}{Akazaki and Hasuo}{2015}]%
        {AkazakiHasuo2015}
\bibfield{author}{\bibinfo{person}{Takumi Akazaki} {and}
  \bibinfo{person}{Ichiro Hasuo}.} \bibinfo{year}{2015}\natexlab{}.
\newblock \showarticletitle{Time robustness in MTL and expressivity in hybrid
  system falsification}. In \bibinfo{booktitle}{\emph{Computer Aided
  Verification (CAV)}} \emph{(\bibinfo{series}{LNCS})},
  Vol.~\bibinfo{volume}{9207}. Springer, \bibinfo{pages}{356--374}.
\newblock


\bibitem[\protect\citeauthoryear{Annpureddy, Liu, Fainekos, and
  Sankaranarayanan}{Annpureddy et~al\mbox{.}}{2011}]%
        {Annpureddy-et-al2011}
\bibfield{author}{\bibinfo{person}{Yashwanth Annpureddy}, \bibinfo{person}{Che
  Liu}, \bibinfo{person}{Georgios Fainekos}, {and} \bibinfo{person}{Sriram
  Sankaranarayanan}.} \bibinfo{year}{2011}\natexlab{}.
\newblock \showarticletitle{S-TaLiRo: A Tool for Temporal Logic Falsification
  for Hybrid Systems}. In \bibinfo{booktitle}{\emph{Tools and Algorithms for
  the Construction and Analysis of Systems (TACAS)}}
  \emph{(\bibinfo{series}{LNCS})},
  \bibfield{editor}{\bibinfo{person}{Parosh~Aziz Abdulla} {and}
  \bibinfo{person}{K.~Rustan~M. Leino}} (Eds.). \bibinfo{publisher}{Springer},
  \bibinfo{pages}{254--257}.
\newblock


\bibitem[\protect\citeauthoryear{Balkan, Tabuada, Deshmukh, Jin, and
  Kapinski}{Balkan et~al\mbox{.}}{2017}]%
        {Underminer}
\bibfield{author}{\bibinfo{person}{Ayca Balkan}, \bibinfo{person}{Paulo
  Tabuada}, \bibinfo{person}{Jyotirmoy~V. Deshmukh}, \bibinfo{person}{Xiaoqing
  Jin}, {and} \bibinfo{person}{James Kapinski}.}
  \bibinfo{year}{2017}\natexlab{}.
\newblock \showarticletitle{Underminer: A Framework for Automatically
  Identifying Nonconverging Behaviors in Black-Box System Models}.
\newblock \bibinfo{journal}{\emph{ACM Trans. Embed. Comput. Syst.}}
  \bibinfo{volume}{17}, \bibinfo{number}{1}, Article \bibinfo{articleno}{20}
  (\bibinfo{year}{2017}), \bibinfo{numpages}{28}~pages.
\newblock


\bibitem[\protect\citeauthoryear{Barrett, Fontaine, and Tinelli}{Barrett
  et~al\mbox{.}}{2017}]%
        {smtlib}
\bibfield{author}{\bibinfo{person}{Clark Barrett}, \bibinfo{person}{Pascal
  Fontaine}, {and} \bibinfo{person}{Cesare Tinelli}.}
  \bibinfo{year}{2017}\natexlab{}.
\newblock \bibinfo{booktitle}{\emph{The {SMT-LIB} Standard: {V}ersion 2.6}}.
\newblock \bibinfo{type}{{T}echnical {R}eport}.
  \bibinfo{institution}{Department of Computer Science, The University of
  Iowa}.
\newblock
\newblock
\shownote{Available at {\tt www.SMT-LIB.org}.}


\bibitem[\protect\citeauthoryear{Bartocci, Deshmukh, Donz{\'e}, Fainekos,
  Maler, Ni{\v{c}}kovi{\'c}, and Sankaranarayanan}{Bartocci
  et~al\mbox{.}}{2018}]%
        {BDDFMNS2018}
\bibfield{author}{\bibinfo{person}{Ezio Bartocci}, \bibinfo{person}{Jyotirmoy
  Deshmukh}, \bibinfo{person}{Alexandre Donz{\'e}}, \bibinfo{person}{Georgios
  Fainekos}, \bibinfo{person}{Oded Maler}, \bibinfo{person}{Dejan
  Ni{\v{c}}kovi{\'c}}, {and} \bibinfo{person}{Sriram Sankaranarayanan}.}
  \bibinfo{year}{2018}\natexlab{}.
\newblock \showarticletitle{Specification-based monitoring of cyber-physical
  systems: a survey on theory, tools and applications}.
\newblock In \bibinfo{booktitle}{\emph{Lectures on Runtime Verification}}.
  \bibinfo{publisher}{Springer}, \bibinfo{pages}{135--175}.
\newblock


\bibitem[\protect\citeauthoryear{Deshmukhand, Jin, Kapinski, and
  Maler}{Deshmukhand et~al\mbox{.}}{[n. d.]}]%
        {DeshmukhJKM2015}
\bibfield{author}{\bibinfo{person}{Jyotirmoy Deshmukhand},
  \bibinfo{person}{Xiaoqing Jin}, \bibinfo{person}{James Kapinski}, {and}
  \bibinfo{person}{Oded Maler}.} \bibinfo{year}{[n. d.]}\natexlab{}.
\newblock \showarticletitle{Stochastic Local Search for Falsification of Hybrid
  Systems}.
\newblock \bibinfo{journal}{\emph{Automated Technology for Verification and
  Analysis (ATVA)}}  \bibinfo{volume}{9364} (\bibinfo{year}{[n. d.]}),
  \bibinfo{pages}{500--517}.
\newblock


\bibitem[\protect\citeauthoryear{Dokhanchi, Yaghoubi, Hoxha, and
  Fainekos}{Dokhanchi et~al\mbox{.}}{2017}]%
        {Dokhanchi-et-al2017}
\bibfield{author}{\bibinfo{person}{Adel Dokhanchi}, \bibinfo{person}{Shakiba
  Yaghoubi}, \bibinfo{person}{Bardh Hoxha}, {and} \bibinfo{person}{Georgios~E.
  Fainekos}.} \bibinfo{year}{2017}\natexlab{}.
\newblock \showarticletitle{{ARCH-COMP17} Category Report: Preliminary Results
  on the Falsification Benchmarks}. In \bibinfo{booktitle}{\emph{Applied
  Verification of Continuous and Hybrid Systems (ARCH)}}
  \emph{(\bibinfo{series}{EPiC Series in Computing})},
  \bibfield{editor}{\bibinfo{person}{Goran Frehse} {and}
  \bibinfo{person}{Matthias Althoff}} (Eds.), Vol.~\bibinfo{volume}{48}.
  \bibinfo{publisher}{EasyChair}, \bibinfo{pages}{170--174}.
\newblock


\bibitem[\protect\citeauthoryear{Dokhanchi, Yaghoubi, Hoxha, Fainekos, Ernst,
  Zhang, Arcaini, Hasuo, and Sedwards}{Dokhanchi et~al\mbox{.}}{2019}]%
        {Dokhanchi-et-al2018}
\bibfield{author}{\bibinfo{person}{Adel Dokhanchi}, \bibinfo{person}{Shakiba
  Yaghoubi}, \bibinfo{person}{Bardh Hoxha}, \bibinfo{person}{Georgios~E.
  Fainekos}, \bibinfo{person}{Gidon Ernst}, \bibinfo{person}{Zhenya Zhang},
  \bibinfo{person}{Paolo Arcaini}, \bibinfo{person}{Ichiro Hasuo}, {and}
  \bibinfo{person}{Sean Sedwards}.} \bibinfo{year}{2019}\natexlab{}.
\newblock \showarticletitle{{ARCH-COMP18} Category Report: Results on the
  Falsification Benchmarks}. In \bibinfo{booktitle}{\emph{Applied Verification
  of Continuous and Hybrid Systems (ARCH)}} \emph{(\bibinfo{series}{EPiC Series
  in Computing})}, \bibfield{editor}{\bibinfo{person}{Goran Frehse}} (Ed.),
  Vol.~\bibinfo{volume}{54}. \bibinfo{publisher}{EasyChair},
  \bibinfo{pages}{104--109}.
\newblock


\bibitem[\protect\citeauthoryear{Donz{\'e}}{Donz{\'e}}{2010}]%
        {Donze2010}
\bibfield{author}{\bibinfo{person}{Alexandre Donz{\'e}}.}
  \bibinfo{year}{2010}\natexlab{}.
\newblock \showarticletitle{{Breach}, {A} toolbox for verification and
  parameter synthesis of hybrid systems}. In
  \bibinfo{booktitle}{\emph{International Conference on Computer Aided
  Verification (CAV 2010)}} \emph{(\bibinfo{series}{LNCS})}.
  \bibinfo{publisher}{Springer}, \bibinfo{pages}{167--170}.
\newblock


\bibitem[\protect\citeauthoryear{Donz{\'{e}}, Ferr{\`{e}}re, and
  Maler}{Donz{\'{e}} et~al\mbox{.}}{2013}]%
        {DonzeFerrereMaler2013}
\bibfield{author}{\bibinfo{person}{Alexandre Donz{\'{e}}},
  \bibinfo{person}{Thomas Ferr{\`{e}}re}, {and} \bibinfo{person}{Oded Maler}.}
  \bibinfo{year}{2013}\natexlab{}.
\newblock \showarticletitle{Efficient Robust Monitoring for {STL}}. In
  \bibinfo{booktitle}{\emph{Computer Aided Verification (CAV)}}
  \emph{(\bibinfo{series}{LNCS})}, \bibfield{editor}{\bibinfo{person}{Natasha
  Sharygina} {and} \bibinfo{person}{Helmut Veith}} (Eds.),
  Vol.~\bibinfo{volume}{8044}. \bibinfo{publisher}{Springer},
  \bibinfo{pages}{264--279}.
\newblock


\bibitem[\protect\citeauthoryear{Donz{\'e} and Maler}{Donz{\'e} and
  Maler}{2010}]%
        {DonzeMaler2010}
\bibfield{author}{\bibinfo{person}{Alexandre Donz{\'e}} {and}
  \bibinfo{person}{Oded Maler}.} \bibinfo{year}{2010}\natexlab{}.
\newblock \showarticletitle{Robust Satisfaction of Temporal Logic over
  Real-Valued Signals}. In \bibinfo{booktitle}{\emph{Formal Modeling and
  Analysis of Timed Systems (FORMATS 2010)}} \emph{(\bibinfo{series}{LNCS})},
  \bibfield{editor}{\bibinfo{person}{Krishnendu Chatterjee} {and}
  \bibinfo{person}{Thomas~A. Henzinger}} (Eds.). \bibinfo{publisher}{Springer},
  \bibinfo{pages}{92--106}.
\newblock


\bibitem[\protect\citeauthoryear{Dreossi, Dang, Donz{\'e}, Kapinski, Jin, and
  Deshmukh}{Dreossi et~al\mbox{.}}{2015}]%
        {Dreossi2015}
\bibfield{author}{\bibinfo{person}{Tommaso Dreossi}, \bibinfo{person}{Thao
  Dang}, \bibinfo{person}{Alexandre Donz{\'e}}, \bibinfo{person}{James
  Kapinski}, \bibinfo{person}{Xiaoqing Jin}, {and}
  \bibinfo{person}{Jyotirmoy~V. Deshmukh}.} \bibinfo{year}{2015}\natexlab{}.
\newblock \showarticletitle{Efficient Guiding Strategies for Testing of
  Temporal Properties of Hybrid Systems}. In \bibinfo{booktitle}{\emph{NASA
  Formal Methods (NFM)}} \emph{(\bibinfo{series}{LNCS})},
  \bibfield{editor}{\bibinfo{person}{Klaus Havelund}, \bibinfo{person}{Gerard
  Holzmann}, {and} \bibinfo{person}{Rajeev Joshi}} (Eds.),
  Vol.~\bibinfo{volume}{9058}. \bibinfo{publisher}{Springer},
  \bibinfo{pages}{127--142}.
\newblock


\bibitem[\protect\citeauthoryear{Eddeland, Miremadi, Fabian, and
  {\AA}kesson}{Eddeland et~al\mbox{.}}{2017}]%
        {EddelandMFA2017}
\bibfield{author}{\bibinfo{person}{Johan Eddeland}, \bibinfo{person}{Sajed
  Miremadi}, \bibinfo{person}{Martin Fabian}, {and} \bibinfo{person}{Knut
  {\AA}kesson}.} \bibinfo{year}{2017}\natexlab{}.
\newblock \showarticletitle{Objective functions for falsification of signal
  temporal logic properties in cyber-physical systems}. In
  \bibinfo{booktitle}{\emph{Conference on Automation Science and Engineering
  (CASE)}}. IEEE, \bibinfo{pages}{1326--1331}.
\newblock


\bibitem[\protect\citeauthoryear{Fainekos and Pappas}{Fainekos and
  Pappas}{2009}]%
        {FainekosPappas2009journal}
\bibfield{author}{\bibinfo{person}{Georgios~E. Fainekos} {and}
  \bibinfo{person}{George~J. Pappas}.} \bibinfo{year}{2009}\natexlab{}.
\newblock \showarticletitle{Robustness of temporal logic specifications for
  continuous-time signals}.
\newblock \bibinfo{journal}{\emph{Theor. Comp. Sci.}} \bibinfo{volume}{410},
  \bibinfo{number}{42} (\bibinfo{year}{2009}), \bibinfo{pages}{4262--4291}.
\newblock


\bibitem[\protect\citeauthoryear{Fleming and Wallace}{Fleming and
  Wallace}{1986}]%
        {FlemingW1986}
\bibfield{author}{\bibinfo{person}{Philip~J Fleming} {and}
  \bibinfo{person}{John~J Wallace}.} \bibinfo{year}{1986}\natexlab{}.
\newblock \showarticletitle{How not to lie with statistics: the correct way to
  summarize benchmark results}.
\newblock \bibinfo{journal}{\emph{Commun. ACM}} \bibinfo{volume}{29},
  \bibinfo{number}{3} (\bibinfo{year}{1986}), \bibinfo{pages}{218--221}.
\newblock


\bibitem[\protect\citeauthoryear{Hoxha, Abbas, and Fainekos}{Hoxha
  et~al\mbox{.}}{2014}]%
        {HoxhaAbbasFainekos2014}
\bibfield{author}{\bibinfo{person}{Bardh Hoxha}, \bibinfo{person}{Houssam
  Abbas}, {and} \bibinfo{person}{Georgios~E. Fainekos}.}
  \bibinfo{year}{2014}\natexlab{}.
\newblock \showarticletitle{Benchmarks for Temporal Logic Requirements for
  Automotive Systems}. In \bibinfo{booktitle}{\emph{Applied veRification for
  Continuous and Hybrid Systems (ARCH)}} \emph{(\bibinfo{series}{EPiC Series in
  Computing})}, \bibfield{editor}{\bibinfo{person}{Goran Frehse} {and}
  \bibinfo{person}{Matthias Althoff}} (Eds.), Vol.~\bibinfo{volume}{34}.
  \bibinfo{publisher}{EasyChair}, \bibinfo{pages}{25--30}.
\newblock


\bibitem[\protect\citeauthoryear{Igel, Hansen, and Roth}{Igel
  et~al\mbox{.}}{2007}]%
        {Igel2007}
\bibfield{author}{\bibinfo{person}{Christian Igel}, \bibinfo{person}{Nikolaus
  Hansen}, {and} \bibinfo{person}{Stefan Roth}.}
  \bibinfo{year}{2007}\natexlab{}.
\newblock \showarticletitle{Covariance matrix adaptation for multi-objective
  optimization}.
\newblock \bibinfo{journal}{\emph{Evolutionary computation}}
  \bibinfo{volume}{15}, \bibinfo{number}{1} (\bibinfo{year}{2007}),
  \bibinfo{pages}{1--28}.
\newblock


\bibitem[\protect\citeauthoryear{Jegourel, Legay, and Sedwards}{Jegourel
  et~al\mbox{.}}{2013}]%
        {JegourelLegaySedwards2013}
\bibfield{author}{\bibinfo{person}{Cyrille Jegourel}, \bibinfo{person}{Axel
  Legay}, {and} \bibinfo{person}{Sean Sedwards}.}
  \bibinfo{year}{2013}\natexlab{}.
\newblock \showarticletitle{Importance Splitting for Statistical Model Checking
  Rare Properties}.
\newblock In \bibinfo{booktitle}{\emph{Computer Aided Verification (CAV)}}.
  \bibinfo{series}{LNCS}, Vol.~\bibinfo{volume}{8044}.
  \bibinfo{publisher}{Springer}, \bibinfo{pages}{576--591}.
\newblock


\bibitem[\protect\citeauthoryear{Jegourel, Legay, and Sedwards}{Jegourel
  et~al\mbox{.}}{2014}]%
        {JegourelLegaySedwards2014}
\bibfield{author}{\bibinfo{person}{Cyrille Jegourel}, \bibinfo{person}{Axel
  Legay}, {and} \bibinfo{person}{Sean Sedwards}.}
  \bibinfo{year}{2014}\natexlab{}.
\newblock \showarticletitle{An Effective Heuristic for Adaptive Importance
  Splitting in Statistical Model Checking}.
\newblock In \bibinfo{booktitle}{\emph{Leveraging Applications of Formal
  Methods, Verification and Validation. Specialized Techniques and Applications
  (ISoLA)}}, \bibfield{editor}{\bibinfo{person}{Tiziana Margaria} {and}
  \bibinfo{person}{Bernhard Steffen}} (Eds.). \bibinfo{series}{LNCS},
  Vol.~\bibinfo{volume}{8803}. \bibinfo{publisher}{Springer},
  \bibinfo{pages}{143--159}.
\newblock


\bibitem[\protect\citeauthoryear{Jin, Deshmukh, Kapinski, Ueda, and Butts}{Jin
  et~al\mbox{.}}{2014}]%
        {Jin-et-al2014}
\bibfield{author}{\bibinfo{person}{Xiaoqing Jin}, \bibinfo{person}{Jyotirmoy~V.
  Deshmukh}, \bibinfo{person}{James Kapinski}, \bibinfo{person}{Koichi Ueda},
  {and} \bibinfo{person}{Kenneth~R. Butts}.} \bibinfo{year}{2014}\natexlab{}.
\newblock \showarticletitle{Powertrain control verification benchmark}. In
  \bibinfo{booktitle}{\emph{Hybrid Systems: Computation and Control (HSCC)}},
  \bibfield{editor}{\bibinfo{person}{Martin Fr{\"{a}}nzle} {and}
  \bibinfo{person}{John Lygeros}} (Eds.). \bibinfo{publisher}{{ACM}},
  \bibinfo{pages}{253--262}.
\newblock


\bibitem[\protect\citeauthoryear{Jin, Donz{\'e}, Deshmukh, and Seshia}{Jin
  et~al\mbox{.}}{2015}]%
        {JinDDS2015}
\bibfield{author}{\bibinfo{person}{Xiaoqing Jin}, \bibinfo{person}{Alexandre
  Donz{\'e}}, \bibinfo{person}{Jyotirmoy~V Deshmukh}, {and}
  \bibinfo{person}{Sanjit~A Seshia}.} \bibinfo{year}{2015}\natexlab{}.
\newblock \showarticletitle{Mining requirements from closed-loop control
  models}.
\newblock \bibinfo{journal}{\emph{IEEE Transactions on Computer-Aided Design of
  Integrated Circuits and Systems}} \bibinfo{volume}{34}, \bibinfo{number}{11}
  (\bibinfo{year}{2015}), \bibinfo{pages}{1704--1717}.
\newblock


\bibitem[\protect\citeauthoryear{Kapinski, Deshmukh, Jin, Ito, and
  Butts}{Kapinski et~al\mbox{.}}{2016}]%
        {KDJIB16}
\bibfield{author}{\bibinfo{person}{James Kapinski},
  \bibinfo{person}{Jyotirmoy~V. Deshmukh}, \bibinfo{person}{Xiaoqing Jin},
  \bibinfo{person}{Hisahiro Ito}, {and} \bibinfo{person}{Ken Butts}.}
  \bibinfo{year}{2016}\natexlab{}.
\newblock \showarticletitle{Simulation-Based Approaches for Verification of
  Embedded Control Systems: An Overview of Traditional and Advanced Modeling,
  Testing, and Verification Techniques}.
\newblock \bibinfo{journal}{\emph{IEEE Control Systems Magazine}}
  \bibinfo{volume}{36}, \bibinfo{number}{6} (\bibinfo{date}{Dec}
  \bibinfo{year}{2016}), \bibinfo{pages}{45--64}.
\newblock


\bibitem[\protect\citeauthoryear{Kocsis and Szepesv{\'a}ri}{Kocsis and
  Szepesv{\'a}ri}{2006}]%
        {KocsisSzepesvari2006}
\bibfield{author}{\bibinfo{person}{Levente Kocsis} {and} \bibinfo{person}{Csaba
  Szepesv{\'a}ri}.} \bibinfo{year}{2006}\natexlab{}.
\newblock \showarticletitle{Bandit Based {M}onte-{C}arlo Planning}. In
  \bibinfo{booktitle}{\emph{Machine Learning (ECML 2006)}}
  \emph{(\bibinfo{series}{LNCS})}, \bibfield{editor}{\bibinfo{person}{Johannes
  F{\"u}rnkranz}, \bibinfo{person}{Tobias Scheffer}, {and}
  \bibinfo{person}{Myra Spiliopoulou}} (Eds.), Vol.~\bibinfo{volume}{4212}.
  \bibinfo{publisher}{Springer}, \bibinfo{pages}{282--293}.
\newblock


\bibitem[\protect\citeauthoryear{LaValle and Kuffner~Jr}{LaValle and
  Kuffner~Jr}{2001}]%
        {LaValleKuffner2001}
\bibfield{author}{\bibinfo{person}{Steven~M LaValle} {and}
  \bibinfo{person}{James~J Kuffner~Jr}.} \bibinfo{year}{2001}\natexlab{}.
\newblock \showarticletitle{Randomized kinodynamic planning}.
\newblock \bibinfo{journal}{\emph{The International Journal of Robotics
  Research (IJRR)}} \bibinfo{volume}{20}, \bibinfo{number}{5}
  (\bibinfo{year}{2001}), \bibinfo{pages}{378--400}.
\newblock


\bibitem[\protect\citeauthoryear{Lee, Kochenderfer, Mengshoel, Brat, and
  Owen}{Lee et~al\mbox{.}}{2015}]%
        {LKMBO15}
\bibfield{author}{\bibinfo{person}{Ritchie Lee}, \bibinfo{person}{Mykel~J.
  Kochenderfer}, \bibinfo{person}{Ole~J. Mengshoel},
  \bibinfo{person}{Guillaume~P. Brat}, {and} \bibinfo{person}{Michael~P.
  Owen}.} \bibinfo{year}{2015}\natexlab{}.
\newblock \showarticletitle{Adaptive stress testing of airborne collision
  avoidance systems}. In \bibinfo{booktitle}{\emph{IEEE/AIAA 34th Digital
  Avionics Systems Conference (DASC 2015)}}. \bibinfo{pages}{6C2:1--6C2:13}.
\newblock
\showISSN{2155-7195}


\bibitem[\protect\citeauthoryear{Sankaranarayanan and
  Fainekos}{Sankaranarayanan and Fainekos}{2012}]%
        {SankaranarayananFainekos2012}
\bibfield{author}{\bibinfo{person}{Sriram Sankaranarayanan} {and}
  \bibinfo{person}{Georgios Fainekos}.} \bibinfo{year}{2012}\natexlab{}.
\newblock \showarticletitle{Falsification of temporal properties of hybrid
  systems using the cross-entropy method}. In \bibinfo{booktitle}{\emph{Hybrid
  Systems: Computation and Control (HSCC)}}. ACM, \bibinfo{pages}{125--134}.
\newblock


\bibitem[\protect\citeauthoryear{Sutton and Barto}{Sutton and Barto}{2018}]%
        {SB2018}
\bibfield{author}{\bibinfo{person}{Richard~S. Sutton} {and}
  \bibinfo{person}{Andrew~G. Barto}.} \bibinfo{year}{2018}\natexlab{}.
\newblock \bibinfo{booktitle}{\emph{Reinforcement Learning: An Introduction}
  (\bibinfo{edition}{second} ed.)}.
\newblock \bibinfo{publisher}{MIT press}.
\newblock


\bibitem[\protect\citeauthoryear{Wolpert and Macready}{Wolpert and
  Macready}{1997}]%
        {WolpertMacready1997}
\bibfield{author}{\bibinfo{person}{David Wolpert} {and}
  \bibinfo{person}{William~G. Macready}.} \bibinfo{year}{1997}\natexlab{}.
\newblock \showarticletitle{No free lunch theorems for optimization}.
\newblock \bibinfo{journal}{\emph{{IEEE} Trans. Evolutionary Computation}}
  \bibinfo{volume}{1}, \bibinfo{number}{1} (\bibinfo{year}{1997}),
  \bibinfo{pages}{67--82}.
\newblock


\bibitem[\protect\citeauthoryear{Zhang, Ernst, Hasuo, and Sedwards}{Zhang
  et~al\mbox{.}}{2018a}]%
        {ZEHS18}
\bibfield{author}{\bibinfo{person}{Zhenya Zhang}, \bibinfo{person}{Gidon
  Ernst}, \bibinfo{person}{Ichiro Hasuo}, {and} \bibinfo{person}{Sean
  Sedwards}.} \bibinfo{year}{2018}\natexlab{a}.
\newblock \showarticletitle{Time-Staging Enhancement of Hybrid System
  Falsification}. In \bibinfo{booktitle}{\emph{2018 IEEE Workshop on Monitoring
  and Testing of Cyber-Physical Systems (MT-CPS 2018)}}.
  \bibinfo{publisher}{IEEE}, \bibinfo{pages}{3--4}.
\newblock


\bibitem[\protect\citeauthoryear{Zhang, Ernst, Sedwards, Arcaini, and
  Hasuo}{Zhang et~al\mbox{.}}{2018b}]%
        {ZESAH18}
\bibfield{author}{\bibinfo{person}{Zhenya Zhang}, \bibinfo{person}{Gidon
  Ernst}, \bibinfo{person}{Sean Sedwards}, \bibinfo{person}{Paolo Arcaini},
  {and} \bibinfo{person}{Ichiro Hasuo}.} \bibinfo{year}{2018}\natexlab{b}.
\newblock \showarticletitle{Two-Layered Falsification of Hybrid Systems Guided
  by {M}onte {C}arlo Tree Search}.
\newblock \bibinfo{journal}{\emph{IEEE Transactions on Computer-Aided Design of
  Integrated Circuits and Systems (TCAD 2018)}} (\bibinfo{year}{2018}).
\newblock


\bibitem[\protect\citeauthoryear{Zutshi, Deshmukh, Sankaranarayanan, and
  Kapinski}{Zutshi et~al\mbox{.}}{2014}]%
        {ZutshiDSK14}
\bibfield{author}{\bibinfo{person}{Aditya Zutshi},
  \bibinfo{person}{Jyotirmoy~V. Deshmukh}, \bibinfo{person}{Sriram
  Sankaranarayanan}, {and} \bibinfo{person}{James Kapinski}.}
  \bibinfo{year}{2014}\natexlab{}.
\newblock \showarticletitle{Multiple shooting, CEGAR-based falsification for
  hybrid systems}. In \bibinfo{booktitle}{\emph{Embedded Software (EMSOFT)}}.
  \bibinfo{pages}{5:1--5:10}.
\newblock


\end{thebibliography}

\end{document}